\documentstyle[aps,prb,twocolumn,epsf,floats]{revtex}
\draft
\begin{document}
\wideabs{

\title{Energy spectra of fractional quantum Hall systems 
       in the presence of a valence hole}

\author{
   Arkadiusz W\'ojs}
\address{
   Department of Physics, 
   University of Tennessee, Knoxville, Tennessee 37996 \\
   Institute of Physics, 
   Wroclaw University of Technology, Wroclaw 50-370, Poland}

\author{
   John J. Quinn}
\address{
   Department of Physics, 
   University of Tennessee, Knoxville, Tennessee 37996}

\maketitle

\begin{abstract}
The energy spectrum of a two-dimensional electron gas (2DEG) in the 
fractional quantum Hall regime interacting with an optically injected 
valence band hole is studied as a function of the filling factor $\nu$ 
and the separation $d$ between the electron and hole layers.
The response of the 2DEG to the hole changes abruptly at $d$ of the 
order of the magnetic length $\lambda$.
At $d<\lambda$, the hole binds electrons to form neutral ($X$) or 
charged ($X^-$) excitons, and the photoluminescence (PL) spectrum 
probes the lifetimes and binding energies of these states rather 
than the original correlations of the 2DEG.
The ``dressed exciton'' picture (in which the interaction between 
an exciton and the 2DEG was proposed to merely enhance the exciton 
mass) is questioned.
Instead, the low energy states are explained in terms of Laughlin 
correlations between the constituent fermions (electrons and $X^-$'s)
and the formation of two-component incompressible fluid states in the 
electron--hole plasma.
At $d>2\lambda$, the hole binds up to two Laughlin quasielectrons 
(QE) of the 2DEG to form fractionally charged excitons $h$QE$_n$.
The previously found ``anyon exciton'' $h$QE$_3$ is shown to be 
unstable at any value of $d$.
The critical dependence of the stability of different $h$QE$_n$ 
complexes on the presence of QE's in the 2DEG leads to the observed 
discontinuity of the PL spectrum at $\nu={1\over3}$ or ${2\over3}$.
\end{abstract}
}

\section{Introduction}
\label{secI}
A number of experimental\cite{heiman,turberfield,goldberg,buhmann1,goldys,%
kukushkin,takeyama,gravier,pinczuk,kheng,buhmann2,shields1,finkelstein,%
hayne,nickel,tischler,wojtowicz,jiang,brown,kim} and theoretical
\cite{lerner,dzyubenko1,macdonald1,macdonald2,wang,apalkov,rashba,chen,%
stebe,x-dot,palacios,x-fqhe,x-cf,x-pl,whittaker,pawel1,pawel2} studies 
of the optical properties of quasi-two-dimensional (2D) electron systems 
in high magnetic fields have been carried out in the recent years.
In structures where both conduction electrons and valence holes are 
confined in the same 2D layer, such as symmetrically doped quantum 
wells (QW's), the photoluminescence (PL) spectrum of an electron gas 
(2DEG) involves neutral and charged exciton complexes (bound states 
of one or two electrons and a hole, $X=e$--$h$ and $X^-=2e$--$h$).
\cite{kheng,buhmann2,shields1,finkelstein,hayne,nickel,tischler,%
wojtowicz,jiang,brown,kim,stebe,x-dot,palacios,x-fqhe,x-cf,x-pl,%
whittaker}
The $X^-$ can exist in the form of a number of different bound states.
In zero or low magnetic field ($B\le2$~T in GaAs), only the optically 
active spin--singlet $X^-_s$ occurs.\cite{stebe,x-pl,whittaker}
Although it is predicted to unbind in the $B\rightarrow\infty$ limit as 
a consequence of the ``hidden symmetry'' of an $e$--$h$ system in the 
lowest LL,\cite{lerner,dzyubenko1,macdonald1} the $X^-_s$ is observed 
in the PL spectra even in the highest fields available experimentally 
($\sim50$~T in GaAs).\cite{hayne}
A different $X^-$ bound state is formed in a finite magnetic field:
a non-radiative (``dark'') spin-triplet $X^-_{td}$.\cite{shields1,%
finkelstein}
In contrast with an earlier prediction,\cite{macdonald1} the $X^-_{td}$ 
remains bound in the $B\rightarrow\infty$ limit,\cite{x-dot,palacios}
and the transition from the $X^-_s$ to the $X^-_{td}$ ground state 
is expected at $B\approx30$~T (in GaAs).\cite{x-pl,whittaker}
At even higher fields, Laughlin incompressible fluid states of strongly 
bound and long-lived $X^-_{td}$ fermionic quasiparticles were predicted. 
\cite{x-fqhe,x-cf}
Very recently, yet another bound $X^-$ state has been discovered
\cite{x-pl} in a strong (but finite) magnetic field: a radiative 
(``bright'') excited spin-triplet $X^-_{tb}$.
The $X^-_{tb}$ has the smallest binding energy but the largest 
oscillator strength of all $X^-$ states, and dominates the PL spectrum 
at very high magnetic fields.\cite{hayne}

The PL spectra of symmetric QW's are not very useful for studying 
the $e$--$e$ correlations in the 2DEG.
In such systems, the 2DEG responds so strongly to the perturbation 
created by an optically injected hole that the original correlations 
are locally (in the vicinity of the hole) completely replaced by the 
$e$--$h$ correlations describing an $X$ or $X^-$ bound state.
The PL spectra containing more information about the properties of
the 2DEG itself are obtained in bi-layer systems, where the spatial 
separation of electrons and holes reduces the effects of $e$--$h$ 
correlations.\cite{macdonald2}
The bi-layer systems are realized experimentally in heterojunctions 
and asymmetrically doped wide QW's, in which a perpendicular electric 
field causes separation of electron and hole 2D layers by a finite 
distance $d$.
Unless $d$ is smaller than the magnetic length $\lambda$, the PL 
spectra of bi-layer systems show no recombination from $X^-$ states.
Instead, they show anomalies\cite{heiman,turberfield,goldberg,buhmann1,%
goldys,kukushkin,takeyama} at the filling factors $\nu={1\over3}$ and 
${2\over3}$, at which Laughlin incompressible fluid states\cite{laughlin} 
are formed in the 2DEG, and the fractional quantum Hall (FQH) effect
\cite{tsui} is observed in transport experiments.

The bi-layer $e$--$h$ system can be viewed as an example of a more 
general one in which the 2DEG with well defined correlations (e.g., 
Laughlin correlations at $\nu={1\over3}$) is perturbed by a potential 
$V_{UD}$ of an additional charge (mobile, in case of a valence hole), 
with controlled characteristic strength (energy scale) $U$ and range 
(length scale) $D$. 
Although the layer separation $d$ is the only adjustable parameter in 
an $e$--$h$ system, larger control over both $U$ and $D$ is possible 
by replacing the hole with a sharp electrode whose potential and 
distance from the 2DEG can be tuned independently, as in a scanning 
tunneling microscope (STM).\cite{binning}
In another similar system, a charged impurity can be located at 
a controlled distance from the 2DEG.\cite{jiang,rezayi1,fox}
The 2DEG has its own characteristic lengths and energies, such as 
the average distance ($\sim\varrho^{-1/2}=\lambda\sqrt{2\pi/\nu}$) 
and Coulomb energy of a pair of nearest electrons, or the energy 
gap $\varepsilon_{\rm QE}+\varepsilon_{\rm QH}$ to create Laughlin 
quasiparticle excitations and the average separation between them.
Therefore, different types of response of the 2DEG to a perturbation 
$V_{UD}$ are expected depending on the relation between $U$ and $D$, 
and the characteristic lengths and energies of the 2DEG.

Although the properties of bi-layer $e$--$h$ systems in the FQH regime 
have been extensively studied in the past,\cite{macdonald1,macdonald2,%
wang,apalkov,rashba,chen} the existing theory is by no means satisfactory.
For example, we argue that the suggestive concept of a ``dressed 
exciton''\cite{wang,apalkov} at small $d$ is not valid, and that the 
``anyon exciton''\cite{rashba} is not the relevant quasiparticle for 
description of the PL spectra at large $d$.
In the present work, the elementary (``true'') quasiparticles (TQP's) 
of the $e$--$h$ system are identified at an arbitrary layer separation 
$d$.
A unified description of the response of the 2DEG to the perturbing 
potential of an optically injected hole is proposed, and a transition
\cite{macdonald2} from an $e$--$h$ correlated (excitonic) to an $e$--$e$ 
correlated (Laughlin) phase at $d\approx1.5\lambda$ is confirmed.
This transition has a pronounced effect on the optical spectra: 
at larger $d$, the discontinuities occur at $\nu={1\over3}$ and 
${2\over3}$ which allow for the optical probing of the Laughlin 
correlations in the 2DEG.

At small layer separations ($d<\lambda$), we show that the lowest 
energy band of $e$--$h$ states does not describe a magnetoexciton 
dispersion,\cite{macdonald1} and that the ``dressed exciton'' model 
proposed by Wang et al.\cite{wang} and by Apalkov and Rashba\cite{apalkov} 
is not valid. 
Instead, the formation of (two-component) incompressible fluid
\cite{x-cf,laughlin,halperin} $e$--$X^-$ states in an $e$--$h$ 
plasma is demonstrated.
The states previously misinterpreted as the dispersion of a ``dressed 
exciton'' with an enhanced mass are shown to contain an $X^-$ interacting 
with a quasihole (QH) of such incompressible fluid.
The list of possible bound states (TQP's) of the system at $d<\lambda$ 
includes the $X$ state, different $X^-$ states, and the $X^-$QH$_n$ 
states in which one or two QH's of the $e$--$X^-$ fluid are bound to 
an $X^-$.
Which of the TQP's occur at the lowest energy depends critically on 
$d$ and $\nu$.

The dependence of the excitation energy gap of the incompressible 
$e$--$X^-$ states on $d$ is also studied.
The enhancement of the gap at small $d>0$ is predicted for some states.
Combining the present result with Ref.~\onlinecite{x-pl} we find that 
Laughlin $e$--$X^-$ correlations, which isolate the $X^-$'s from the 
surrounding 2DEG, survive (or are enhanced) at small $d$ for all of 
$X^-_s$, $X^-_{td}$, and $X^-_{tb}$ states.
Hence, the understanding of the PL spectra in terms of weakly perturbed 
$X^-$ states remains valid at $d<\lambda$.

At large layer separations ($d>2\lambda$), following the work of 
Chen and Quinn,\cite{chen} we study the formation and properties 
of fractionally charged excitons (FCX's), or ``anyonic ions,'' 
$h$QE$_n$ consisting of $n$ Laughlin quasielectrons (QE) of the 
2DEG bound to a distant hole.
We give a detailed analysis of all FCX complexes in terms of their 
angular momenta and binding energies.
The pseudopotentials\cite{haldane1,parentage} (pair energy as a 
function of pair angular momentum) describing interactions between 
the hole, electrons, and the Laughlin quasiparticles are calculated.
Using the knowledge of the involved interactions we predict the 
stability of $h$QE and $h$QE$_2$ complexes and explain the behavior 
of their binding energy on the layer separation $d$.
Somewhat surprisingly, the $h$QE$_3$ complex is found unstable at 
any value of $d$.

The general analysis sketched above is illustrated with the energy 
spectra obtained in large-scale numerical diagonalization of finite 
systems on a Haldane sphere.\cite{haldane2,wu}
Using Lanczos-based algorithms\cite{lanczos} we were able to calculate 
the exact spectra of up to nine electrons and a hole at the filling 
factors $\nu\approx{1\over3}$.
Since our numerical results obtained for fairly large systems can serve 
as raw ``experimental'' input for further theories, we discuss them in 
some detail the last section.
They agree with all our predictions made throughout the paper.

Although in the present work we study a very ideal $e$--$h$ system 
in the lowest LL, our most important conclusions are qualitative and 
thus apply without change to realistic systems.
To obtain a better quantitative agreement with particular experiments, 
the effects due to the LL mixing (less important at $d\ge2\lambda$) 
and finite QW widths must be included in a standard way (see, e.g., 
Ref.~\onlinecite{x-pl} for $d=0$).
Some of our conclusions should also shed light on the physics of other 
related systems, such as the STM. 
In particular, the screening of a potential of a sharp electrode by 
a 2DEG is expected to involve ``real'' electrons when $U$ is large 
and $D$ is small, and Laughlin quasiparticles in the opposite case.
An asymmetry between the response of a 2DEG to a positively and 
negatively charged electrode is expected in the latter case, because
of very different QE--QE and QH--QH interactions at short range.
Let us also add that the problem at $\nu={2\over3}$ is equivalent 
to that at $\nu={1\over3}$ because of the charge-conjugation symmetry 
in the lowest LL.

The presented identification of bound states ($X$, $X^-$, $X^-$QH$_n$, 
and $h$QE$_n$) in $e$--$h$ systems at an arbitrary $d$ and the study of 
their mutual interactions is necessary for the correct description of the 
PL from the 2DEG in the FQH regime.
While the complete discussion of the optical properties of all bound 
$e$--$h$ states will be presented in a following publication,
\cite{bilayer-pl} let us mention that the translational invariance of 
a 2DEG results in strict optical selection rules for bound states 
(analogous to those forbidding emission from an isolated $X^-_{td}$
\cite{palacios,x-fqhe,x-cf}).
As a result, $h$ (the ``uncorrelated hole'' state), $h$QE* (an excited 
state of an $h$--QE pair), and $h$QE$_2$ are the only stable radiative 
states at large $d$, while the recombination of $h$QE (the ground state 
of an $h$--QE pair) or of (unstable) $h$QE$_3$ is forbidden.
Different optical properties of different $h$QE$_n$ complexes and the 
critical dependence of their stability on the presence of QE's in the 
2DEG explain the discontinuities observed\cite{heiman,turberfield,%
goldberg,buhmann1,goldys,kukushkin,takeyama} in the PL at $\nu={1\over3}$ 
or ${2\over3}$. 

\section{Model}
\label{secII}
We consider a system in which a 2DEG in a strong magnetic field $B$ 
fills a fraction $\nu<1$ of the lowest LL of a narrow QW.
A dilute 2D gas of valence-band holes ($\nu_h\ll\nu$) is confined
to a parallel layer, separated from the electron one by a distance $d$. 
The widths of electron and hole layers are set to zero (finite widths 
can be included through appropriate form-factors reducing the effective 
2D interaction matrix elements\cite{x-pl}), and the mixing with excited 
electron and hole LL's is neglected.
The single-particle states $\left|m\right>$ in the lowest LL are labeled 
by orbital angular momentum, $m=0$, $-1$, $-2$, \dots\ for the electrons 
and $m_h=-m=0$, 1, 2, \dots\ for the holes.
Since $\nu_h\ll\nu$ and no bound complexes involving more than one hole 
(such as biexcitons $X_2=2e$--$2h$) occur at large $B$, the $h$--$h$ 
correlations can be neglected and it is enough to study the interaction 
of the 2DEG with only one hole.
The many-electron--one-hole Hamiltonian can be written as
\begin{equation}
\label{eq1}
  H = \sum_{ijkl} \left( c_i^\dagger c_j^\dagger c_k c_l V^{ee}_{ijkl}
                + c_i^\dagger h_j^\dagger h_k c_l V^{eh}_{ijkl} \right),
\end{equation}
where $c_m^\dagger$ ($h_m^\dagger$) and $c_m$ ($h_m$) create and 
annihilate an electron (hole) in state $\left|m\right>$.
Because of the lowest LL degeneracy, $H$ includes only the $e$--$e$ 
and $e$--$h$ interactions whose two-body matrix elements $V^{ee}$ and 
$V^{eh}$ are defined by the intra- and inter-layer Coulomb potentials,
$V_{ee}(r)=e^2/r$ and $V_{eh}(r)=-e^2/\sqrt{r^2+d^2}$.
The convenient units for length and energy are the magnetic length 
$\lambda$ and the energy $e^2/\lambda$, respectively.
At $d=0$, the $e$--$h$ matrix elements are equal to the $e$--$e$ 
exchange ones, $V^{eh}_{ijkl}=-V^{ee}_{ikjl}$, due to the particle--hole 
symmetry, and at $d>0$ the $e$--$h$ attraction is weaker than the 
$e$--$e$ repulsion (at short range).

The 2D translational invariance of $H$ results in conservation of two 
orbital quantum numbers: the projection of total angular momentum 
${\cal M}=\sum_m(c_m^\dagger c_m-h_m^\dagger h_m)m$ and an additional 
angular momentum quantum number ${\cal K}$ associated with partial 
decoupling of the center-of-mass motion of an $e$--$h$ system in a 
homogeneous magnetic field.\cite{avron,dzyubenko2}
For a system with a finite total charge, ${\cal Q}=\sum_m(h_m^\dagger 
h_m-c_m^\dagger c_m)e\ne0$, the partial decoupling of the center-of-mass 
motion means that the energy spectrum consists of degenerate LL's.
\cite{avron} 
The states within each LL are labeled by ${\cal K}=0$, 1, 2, \dots\ 
and all have the same value of ${\cal L}={\cal M}+{\cal K}$.
Since both ${\cal M}$ and ${\cal K}$ (and hence also ${\cal L}$) commute 
with the PL operator ${\cal P}$, which annihilates an optically active 
(zero-momentum, $k=0$) $e$--$h$ pair (exciton), ${\cal M}$, ${\cal K}$, 
and ${\cal L}$ are all simultaneously conserved in the PL process.

The effects associated with finite (short) range correlations (such 
as formation and properties of bound states) can be studied in finite 
systems by exact numerical diagonalization, provided that the system 
size $R$ can be made larger than the characteristic correlation length 
$\delta$ (i.e. the size of the bound state).
Numerical diagonalization of $H$ for finite numbers of electrons 
($N=\sum_m c_m^\dagger c_m$) and holes ($N_h=\sum_m h_m^\dagger h_m$) 
in a finite physical space (area) requires restriction of single-particle 
electron and hole Hilbert spaces to a finite size.
In the planar geometry, inclusion of only a finite number of electron 
and hole states in the calculation (states with $m$ only up to certain 
value $m_{\rm max}$) breaks the translational symmetry and the 
conservation of ${\cal K}$.
A finite dispersion of calculated LL's, which disappears only in the 
$m_{\rm max}\rightarrow\infty$ limit, hides the underlying symmetry 
of the modeled (infinite) system.
Also, the calculated PL oscillator strengths do not obey the exact
$\Delta{\cal K}=0$ optical selection rule that holds in an infinite 
system.

More informative finite-size spectra are obtained here using Haldane's 
geometry,\cite{haldane2} where electrons and holes are confined to 
a spherical surface of radius $R$ and the radial magnetic field is 
produced by a Dirac monopole.
The reason for choosing the spherical geometry for the calculations 
is strictly technical and of no physical consequence for the results.
Finite area (and thus finite LL degeneracy) of a closed surface results 
in finite size of the many-body Hilbert spaces obtained without breaking 
the 2D translational symmetry of a plane (which is preserved in the form 
of the 2D group of rotations).
The exact mapping\cite{x-pl,geometry} between quantum numbers 
${\cal M}$ and ${\cal K}$ on a plane and the 2D algebra of the total 
angular momentum ${\bf L}$ on a sphere allows investigation of effects 
caused by those symmetries (such as LL degeneracies and optical 
selection rules) and conversion of the numerical results back to 
the planar geometry.
The price paid for closing the Hilbert space without breaking the 
symmetries is the surface curvature which modifies the interaction 
matrix elements $V^{ee}_{ijkl}$ and $V^{eh}_{ijkl}$.
However, if the correlation length $\delta$ can be made smaller than 
$R$ (as happens for both Laughlin correlations in FQH systems and for 
bound states), the effects of curvature are scaled by a small parameter 
$\delta/R$ and can be eliminated by extrapolation of the results to 
$R\rightarrow\infty$ (in a similar way as the results obtained in the
planar geometry can be extrapolated to $m_{\rm max}\rightarrow\infty$).

The detailed description of the Haldane sphere model can be found 
for example in Refs.~\onlinecite{haldane2,wu,fano} (see also 
Refs.~\onlinecite{x-fqhe,x-cf,x-pl} for application to $e$--$h$ 
systems) and will not be repeated here.
The strength $2S$ of the magnetic monopole is defined in the units 
of flux quantum $\phi_0=hc/e$, so that $4\pi R^2B=2S\phi_0$ and the 
magnetic length is $\lambda=R/\sqrt{S}$.
The single-particle states are the eigenstates of angular momentum 
$l\ge S$ and its projection $m$, and are called monopole harmonics.
The single-particle energies fall into $(2l+1)$-fold degenerate angular 
momentum shells (LL's).
The lowest shell has $l=S$ and thus $2S$ is a measure of the system 
size through the LL degeneracy.
The charged many-body $e$--$h$ states form degenerate total angular 
momentum ($L$) multiplets (LL's) of their own.
The total angular momentum projection $L_z$ labels different states of 
the same multiplet just as ${\cal K}$ or ${\cal M}$ did for different 
states of the same LL on a plane.
Different multiplets are labeled by $L$ just as different LL's on 
a plane were labeled by ${\cal L}$.
The pair of optical selection rules, $\Delta L_z=\Delta L=0$ (equivalent 
to $\Delta{\cal M}=\Delta{\cal K}=0$ on a plane) results from the fact 
that an optically active exciton carries no angular momentum, $l_X=0$.

It is clear that certain properties of a ``strictly'' spherical system 
do not describe the infinite planar system that we intend to model.
For example, if understood literally, finite separation $d$ between 
the electron and hole spheres would lead to different values of the 
magnetic length in the two layers, and thus introduce an asymmetry 
between electron and hole orbitals (even in the lowest LL).
While this effect disappears in the $R\rightarrow\infty$ limit, it 
is eliminated by formally calculating the matrix elements of the 
interaction potential $V_{eh}(r)$ at any value of $d$ for electrons 
and holes confined to a sphere of the same radius $R$.
This procedure justifies the use of spherical geometry at arbitrarily 
large layer separation (not only at $d\ll R$).

\section{Bound Electron--Hole States in a Dilute 2DEG}
\label{secIII}
In order to understand PL from a 2DEG at arbitrary filling factor $\nu$ 
and layer separation $d$, one must first identify the bound complexes 
in which the holes (minority charges) can occur.
After these bound complexes are found and understood in terms of 
such single-particle quantities as total charge ${\cal Q}$, binding 
energy $\Delta$, angular momentum $l$, or PL oscillator strength 
(inverse optical lifetime) $\tau^{-1}$, a perturbation-type analysis 
can be used to determine if those complexes are the relevant 
(or ``true'') quasiparticles (TQP's) of a particular $e$--$h$ system, 
weakly perturbed by interaction with one another and the surrounding 
2DEG.
If it is so, the low energy states can be understood in terms of 
these TQP's and their interactions.
The PL (emission) probes the electron system in the vicinity of the 
annihilated hole and therefore the optical properties of TQP's determine 
the (low temperature) PL spectra of the system.

This type of analysis has been recently applied to the $e$--$h$ 
systems at $d=0$ in the lowest LL,\cite{x-fqhe,x-cf} and it showed 
that the low lying states contained all possible combinations of 
bound $e$--$h$ complexes (excitons $X=e$--$h$ and excitonic ions 
$X_n^-=nX$--$e$) and excess electrons, interacting through effective 
pseudopotentials.
The short range of these pseudopotentials yields Laughlin correlations 
between electrons and excitonic ions, which isolate the latter from the 
2DEG and make them act like well defined TQP's without internal dynamics.
When applied to realistic symmetrically doped ($d=0$) QW's at large $B$
and low density ($\nu<{1\over3}$), a similar analysis showed\cite{x-pl} 
that the observed PL spectra contain transitions only from radiative 
bound states (in that case, spin-singlet and {\em excited} spin-triplet 
$X^-$ states) and explained why the expected\cite{whittaker} 
singlet-triplet $X^-$ crossing was not observed in some experiments.
\cite{hayne}

\subsection{Hidden Symmetry at Zero Layer Separation}
\label{secIII-1}
The exact particle--hole symmetry between electrons and valence holes 
in the lowest LL at $d=0$ results from 
(i) the identical electron and hole single-particle orbitals, scaled 
by the same characteristic length $\lambda$, which yields equal strength 
of $e$--$e$ and $e$--$h$ interaction matrix elements, $V^{eh}_{ijkl}=
-V^{ee}_{ikjl}$, and
(ii) no effects of different effective masses on scattering because 
of the infinite cyclotron gap.
This ``hidden symmetry'' results\cite{macdonald1} in the following 
commutation relation between the Hamiltonian (\ref{eq1}) and the PL 
operator ${\cal P}^\dagger$ which creates a $k=0$ exciton,
\begin{equation}
\label{eq2a}
   [H,{\cal P}^\dagger]=E_X{\cal P}^\dagger,
\end{equation}
where $E_X=-\sqrt{\pi/2}\,e^2/\lambda$ is the exciton energy in the 
lowest LL and ${\cal P}^\dagger=\sum_m(-1)^m c_m^\dagger h_m^\dagger$ 
(on the Haldane's sphere).
Because of Eq.~(\ref{eq2a}), a ``multiplicative'' (MP) eigenstate of 
$H$ (a state containing $N_X$ neutral excitons with momentum zero)
can be constructed by application of ${\cal P}^\dagger$ $N_X$ times
to any eigenstate of the interacting electrons.
The excitons created or annihilated with operators ${\cal P}^\dagger$
and ${\cal P}$ (i.e. by absorption or emission of a photon) have the
same energy $E_X$ which is independent of other electrons or holes 
present.
The number $N_X$ of such ``decoupled'' excitons is conserved by $H$,
only the states with $N_X>0$ are radiative, and the emission (absorption) 
governed by the selection rule $\Delta N_X=-1$ ($+1$) occurs at the bare 
exciton energy $E_X$.\cite{macdonald1}

Somewhat surprisingly, it turns out\cite{chen,x-dot} that the ``totally 
multiplicative'' eigenstate 
\begin{equation}
\label{eq2b}
   \left|\Psi_{N_h}\right>=({\cal P}^\dagger)^{N_h}\left|\Psi\right>,
\end{equation}
obtained by adding the Bose-condensed ground state of $N_X=N_h$ excitons 
each with $k=0$ to the ground state $\left|\Psi\right>$ of excess $N-N_h$ 
electrons, is {\em not always} the ground state of the combined $e$--$h$ 
system.
This results because the interaction of an excited excitonic state
(i.e. one with $k\ne0$) of the Bose condensate with the fluid of 
excess electrons can lower the total energy by more than the cost 
of creating the excited excitonic state.
Typically, a MP state ${\cal P}^\dagger\left|\Phi\right>$ created by 
optical injection of a $k=0$ exciton into a state $\left|\Phi\right>$ 
is an excited state, and the absorption is followed by relaxation to 
a different (non-MP, i.e. non-radiative) ground state.

The condition under which the totally MP state in Eq.~(\ref{eq2b}) 
is the $e$--$h$ ground state follows from the mapping\cite{macdonald1} 
onto the $\uparrow$--$\downarrow$ (spin-unpolarized electron) system, 
in which $\left|\Psi_{N_h}\right>$ corresponds to the 
$\uparrow$--$\downarrow$ state with the maximum spin.
Since $\nu_\uparrow=1-\nu$ and $\nu_\downarrow=\nu_h$, and the 2DEG 
is spin-polarized (in the absence of the Zeeman splitting) only at 
the Laughlin fillings,\cite{rezayi2} the condition for the totally 
MP $e$--$h$ ground state $\left|\Psi_{N_h}\right>$ is
\begin{equation}
\label{eq3}
   \nu-\nu_h=1-(2p+1)^{-1},
\end{equation}
with $p=1$, 2, \dots.
At all other fillings (e.g., $\nu-\nu_h={1\over3}$), the ground state 
has $N_X<N_h$, i.e. contains a number of holes that are bound in other 
(non-radiative) complexes than $k=0$ excitons.

\subsection{Charged Exciton States}
\label{secIII-2}
\begin{figure}[t]
\epsfxsize=3.40in
\epsffile{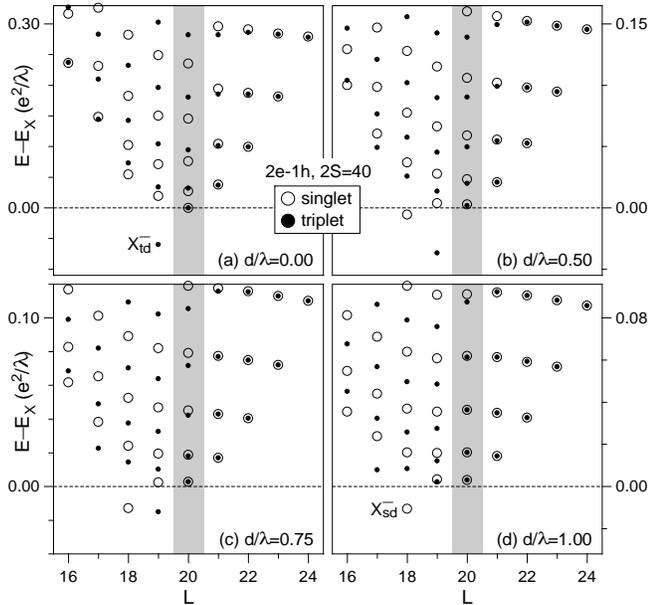}
\caption{
   The energy spectra (energy $E$ vs.\ angular momentum $L$) of 
   the $2e$--$h$ system on a Haldane sphere with the Landau level 
   degeneracy of $2S+1=41$, for different values of the layer 
   separation $d$.
   The open and full circles distinguish states with singlet and 
   triplet electron spin configurations.
   $E_X$ is the exciton energy and $\lambda$ is the magnetic length. 
}
\label{fig01}
\end{figure}

An example of a non-MP $e$--$h$ ground state is the ``dark'' 
spin-triplet charge exciton ($X^-_{td}$).\cite{x-dot}
The $X^-_{td}$ is the only bound $2e$--$h$ state in the lowest LL 
at $d=0$.
It is the most stable $e$--$h$ complex at $\nu_h\le2\nu$, but its 
binding energy decreases at $d>0$ when the $e$--$h$ attraction (at 
short range) becomes smaller than the $e$--$e$ repulsion.
The dependence of the $2e$--$h$ energy spectrum on $d$ is shown in
Fig.~\ref{fig01}.
The spectra are calculated in the spherical geometry for the LL 
degeneracy of $2S+1=41$.
The energy is measured from the exciton energy $E_X$, so that for the 
bound states (the states below the dashed lines) it is the negative of 
the $X^-$ binding energy, $\Delta_{X^-}=E_X-E$.
Open and full symbols distinguish singlet and triplet electron spin 
configurations, and each state with $L>0$ represents a degenerate 
multiplet with $|L_z|\le L$.
The Zeeman energy of the singlet states is not included.
The angular momentum $L$ calculated on a sphere translates into 
the angular momentum quantum numbers on a plane in such way
\cite{x-pl,geometry} that each LL at ${\cal L}=0$, $-1$, $-2$, 
\dots\ (containing states with ${\cal K}=0$, 1, 2, \dots, i.e. with 
${\cal M}={\cal L}-{\cal K}={\cal L}$, ${\cal L}-1$, ${\cal L}-2$, 
\dots) is represented by a multiplet at $L=S+{\cal L}$.
Thus, the low energy multiplets in Fig.~\ref{fig01} at $L=20$, 19, and 
18 represent the planar LL's at ${\cal M}\le{\cal L}=0$, ${\cal M}\le
{\cal L}=-1$, and ${\cal M}\le{\cal L}=-2$, respectively.

It is important to realize that the recombination of an isolated 
$X^-_{td}$ at $d=0$ is forbidden because of two independent symmetries.
\cite{x-fqhe,x-cf,x-pl}
The $\Delta N_X=-1$ selection rule resulting from the hidden symmetry, 
which allows recombination from a pair of MP states at $L=S$ and 
$E=E_X$ only, is lifted at $d>0$.
However, the translational symmetry yielding conservation of $L$ and 
$L_z$ (on a plane, ${\cal M}$ and ${\cal K}$) holds at any value of $d$.
Because the electron left in the lowest LL after recombination has $l=S$ 
(${\cal L}=0$), only those $2e$--$h$ multiplets at $L=S$ (${\cal L}=0$) 
are radiative.
They are marked with shaded rectangles in all frames of Fig.~\ref{fig01}.
In larger systems containing more than a single $X^-$, the translational 
symmetry is broken by collisions, and weak $X^-_{td}$ recombination 
becomes possible.

The $X^-_{td}$ binding energy $\Delta_{X^-_{td}}$, calculated by 
extrapolation of data obtained for $2S\le60$, is about $0.052\,e^2/
\lambda$ at $d=0$ (very close to the value obtained earlier by Palacios 
et al.\cite{palacios} in the planar geometry).
As expected, $\Delta_{X^-_{td}}$ decreases with increasing separation 
up to $d\approx\lambda$, when $X^-_{td}$ unbinds.
Somewhat surprisingly, a new bound multiplet, a singlet $X^-_{sd}$ 
at $L=S-2$ (${\cal L}=-2$), occurs at finite $d$.
Its binding $\Delta_{X^-_{sd}}$ reaches maximum of about $0.013\,
e^2/\lambda$ at $d\approx0.8\lambda$.
The $X^-_{sd}$ is a non-radiative (``dark'') state and should be 
distinguished from the radiative singlet state $X^-_s$ at $L=S$ 
(${\cal L}=0$), which is the $X^-$ ground state at low magnetic 
fields (and small $d$).
The $X^-_{sd}$ is a $2e$--$h$ analog of the singlet $D^-$ state 
(two electrons bound to a distant donor impurity) with the same 
${\cal L}=-2$.
A series of transitions between singlet and triplet $D^-$ states with 
increasing $|{\cal L}|$ have been found when the distance between the 
impurity and the electron layer were increased.\cite{fox}

\begin{figure}[t]
\epsfxsize=3.40in
\epsffile{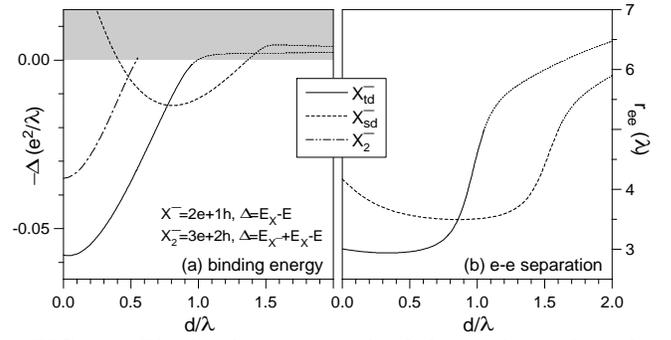}
\caption{
   The binding energy $\Delta$ of the triplet and singlet charged 
   exciton states, $X^-_{td}$ and $X^-_{sd}$, and of the charged 
   biexciton, $X_2^-$, as a function of layer separation $d$.
   $E_X$ is the exciton energy and $\lambda$ is the magnetic length. 
}
\label{fig02}
\end{figure}

Bound states of larger excitonic ions $X_n^-=nX+e$ are also possible 
at small $d$.
They all have completely polarized electron and hole spins, and their 
binding energy, $\Delta_{X_n^-}=E_X+E_{X_{n-1}^-}-E$, decreases with 
increasing size ($n$).
The dependence of $X^-_{td}$, $X^-_{sd}$, and $X_2^-$ binding energies 
(calculated at $2S=60$) on separation $d$ is shown in Fig.~\ref{fig02}(a).
As it was discussed in Sec.~\ref{secII}, finite-size calculations 
give good approximation to $2e$--$h$ energies only for the bound 
(finite-size) states.
While the binding energies are correct at the values of $d$ for which 
$\Delta>0$, they should asymptotically approach zero for $d\rightarrow
\infty$ instead of crossing it as in Fig.~\ref{fig02}(a).
The average $e$--$e$ distance $r_{ee}=\sqrt{\left<{\bf r}_{ee}^2\right>}$ 
within the $X^-_{td}$ and $X^-_{sd}$ complexes is plotted in 
Fig.~\ref{fig02}(b).
Both $X^-$ wavefunctions depend rather weakly on $d$ in the range where 
$\Delta>0$ (i.e., $d\le0.7\lambda$ for $X^-_{td}$ and $0.4\lambda\le 
d\le1.2\lambda$ for $X^-_{sd}$), but when $d$ exceeds the critical 
value ($d=0.8\lambda$ for $X^-_{td}$ and $d=1.3\lambda$ for $X^-_{sd}$), 
$r_{ee}$ quickly increases and the $X^-$ unbinds into an exciton and 
an electron.
Similarly as for binding energies in Fig.~\ref{fig02}(a), we expect 
the $r_{ee}$ curves in Fig.~\ref{fig02}(b) to correctly describe the 
$X^-_{td}$ and $X^-_{sd}$ states on an infinite plane only when 
$r_{ee}$ is smaller than $R\approx5\lambda$.

\section{Electron--Hole States at Small Layer Separation:
         Electron--Charged-Exciton Fluid}
\label{secIV}
\subsection{Zero Layer Separation}
\label{secIV-1}
In the following the 2DEG is assumed to be completely spin-polarized 
because of large Zeeman splitting.
We do not discuss effects due to $X^-_{sd}$ and omit the spin subscript 
in the triplet charged-exciton state $X^-_{td}$.
It follows from Figs.~\ref{fig01} and \ref{fig02} that $X^-$ is the only 
spin-polarized bound $2e$--$h$ state at $d\le\lambda$. 
Since $\Delta_{X^-}>\Delta_{X_2^-}>\Delta_{X_3^-}>\dots$ in entire 
range of $d$, the excitonic ions larger than $X^-$ are unstable in 
the presence of excess electrons (e.g., $X_2^-+e\rightarrow2X^-$), 
and the low lying states at $d<\lambda$ and $\nu_h\ll\nu$ contain 
only $X^-$'s and electrons interacting with one another through 
effective pseudopotentials.\cite{x-fqhe,x-cf}.
The pseudopotential $V_{eX^-}(L)$ (the $e$--$X^-$ pair interaction 
energy $V$ as a function of pair angular momentum $L$) at $d=0$ 
was shown\cite{x-cf} to satisfy the ``short range'' criterion
\cite{parentage} at those values of $L$ which correspond to odd 
``relative'' pair angular momenta ${\cal R}=l_e+l_{X^-}-L$ 
(${\cal R}$ is equal to the usual relative pair angular momentum 
$m$ on a plane).
As a result, generalized Laughlin correlations described in 
the wavefunction by a Jastrow prefactor 
$\prod_{ij}(z_e^{(i)}-z_{X^-}^{(j)})^{m_{eX^-}}$ 
with even exponents $m_{eX^-}$ occur in the two-component 
$e$--$X^-$ fluid.
At certain values of the electron and hole filling factor, these 
correlations result in incompressibility.
For example, the $[m_{ee}m_{X^-X^-}m_{eX^-}]=[332]$ ground state, first 
suggested by Halperin\cite{halperin} for the $\uparrow$--$\downarrow$ 
spin fluid, has been found numerically in the $8e$--$2h$ system.
\cite{x-cf}
A generalized (multi-component) mean-field composite fermion (CF) 
model has been proposed\cite{x-cf} to determine the bands of lowest 
energy states at any $\nu$ and $\nu_h$.
In this model, effective CF magnetic fields of different type (color) 
result which cannot be understood literally.
Rather, the model relies on two simple facts:\cite{x-cf,parentage}
(i) in the low energy states of Laughlin-correlated many-body systems, 
a number of strongly repulsive pair states at the smallest ${\cal R}$ 
are avoided for each type of pair (here, $e$--$e$ and $e$--$X^-$);
(ii) the states satisfying the above constraint can be found more 
easily by noticing that the avoiding of pair states with the smallest
${\cal R}$ is equivalent to the binding of zeros of the many-body 
wavefunction (vortices), which can be reproduced (for the purpose 
of multiplet counting) by attachment of magnetic fluxes.

\begin{figure}[t]
\epsfxsize=3.40in
\epsffile{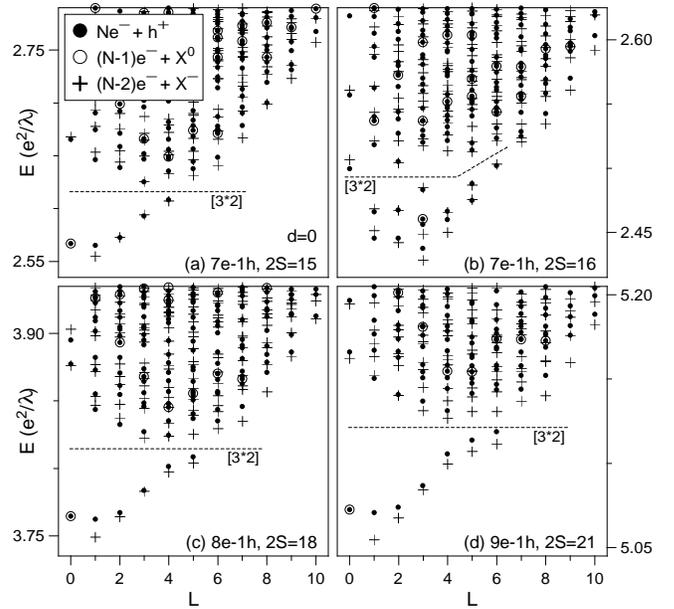}
\caption{
   The energy spectra (energy $E$ vs.\ angular momentum $L$) of the 
   co-planar ($d=0$) $Ne$--$h$ systems on a Haldane sphere with the 
   Landau level degeneracy of $2S+1$: (a) $N=7$ and $2S=15$, (b) $N=7$ 
   and $2S=16$, (c) $N=8$ and $2S=18$, and (d) $N=9$ and $2S=21$.
   Full dots: exact $Ne$--$h$ spectra; open circles: multiplicative
   states; pluses: approximate energies of $(N-2)e$--$X^-$ states.
   The non-multiplicative states below the dashed lines are 
   $(N-2)e$--$X^-$ states with Laughlin--Halperin [3*2] correlations.
   $\lambda$ is the magnetic length. 
}
\label{fig03}
\end{figure}

Let us apply the CF model to the system containing $N$ electrons and 
only one hole.
While the correct picture of this simple system is essential for 
understanding the nature of low energy states and (low-temperature) 
PL of a 2DEG in the FQH regime, it has been interpreted incorrectly 
in a number of previous studies.\cite{rashba}
In Fig.~\ref{fig03} we show the energy spectra for $N=7$, 8, and 9 and 
$2S$ corresponding to $\nu\approx{1\over3}$.
The full dots mark the multiplets obtained in the exact diagonalization
of the $Ne$--$h$ system and the open circles mark the MP states (with 
an $l_X=0$ exciton decoupled from the $N-1$ electron fluid).

In Fig.~\ref{fig03}(acd), the $N-1$ electrons in the lowest energy MP 
state at $L=0$ form the Laughlin $\nu={1\over3}$ ground state.
In Fig.~\ref{fig03}(b), there is one Laughlin quasihole in the lowest 
MP state at $L=3$.
The non-MP low energy states in all frames contain an $X^-$ with 
angular momentum $l_{X^-}=S-1$ and $N-2$ electrons each with $l_e=S$.
The CF picture in which two magnetic fluxes are attached to each 
particle to model the avoiding of the ${\cal R}_{ee}\le2$ and 
${\cal R}_{eX^-}\le1$ pair states yields effective angular momenta 
of $l_e^*=l_e-(N-2)$ and $l_{X^-}^*=l_e^*-1$.
In Fig.~\ref{fig03}(acd) the $N-2$ electrons leave one Laughlin quasihole 
(QH$_e$) with angular momentum $l_{{\rm QH}_e}=l_e^*$ in their $(2l_e^*+
1)$-fold degenerate CF level, and the $X^-$ becomes a single Laughlin 
``quasielectron'' (QE$_{X^-}$) with $l_{{\rm QE}_{X^-}}=l_{X^-}^*$.
The allowed angular momenta $L$ of the QH$_e$--QE$_{X^-}$ pair in the 
lowest energy states of these $(N-2)e$--$X^-$ systems are obtained by 
adding $l_{{\rm QE}_e}$ and $l_{{\rm QE}_{X^-}}$ of two distinguishable 
particles.
The result is: $L=1$, 2, \dots, $N-3$.
Indeed, the multiplets at these values of $L$ form the lowest band 
of non-MP states in Fig.~\ref{fig03}(acd), separated from higher 
states by dashed lines.
The dependence of energy on $L$ within these bands can be interpreted 
as the QH$_e$--QE$_{X^-}$ pseudopotential, and its increase with $L$
means that it is attractive (for a pair of opposite charges, $L$ 
increases with increasing average separation).
The $L=1$ ground states in Fig.~\ref{fig03}(acd) should be therefore
understood as an excitonic bound states of a QH$_e$--QE$_{X^-}$ pair 
in the Laughlin $e$--$X^-$ fluid.
In this state, a Laughlin QH type excitation of charge $+{1\over3}e$ 
is bound to the $X^-$, and the total charge of the $X^-$QH state is
${\cal Q}=-{2\over3}$.
A similar analysis for Fig.~\ref{fig03}(b) gives $l_e^*=3$ and 
$l_{X^-}^*=2$, yielding two QH$_e$'s each with $l_{{\rm QH}_e}=3$ 
and one QE$_{X^-}$ with $l_{{\rm QE}_{X^-}}=2$.
The allowed values of $L$ for such three particles are: $1^2$, $2^2$, 
$3^3$, $4^2$, $5^2$, 6, and 7, exactly as found for the lowest non-MP 
states in Fig.~\ref{fig03}(b).

The strongest indication that the lowest energy bands of non-MP states 
in Fig.~\ref{fig03} contain an $X^-$ interacting with excess electrons 
comes from direct comparison of exact $Ne$--$h$ energies (dots) with 
the approximate energies of the $(N-2)e$--$X^-$ charge configuration 
(pluses). 
The $(N-2)e$--$X^-$ energies are calculated using an effective 
$e$--$X^-$ pseudopotential and the $X^-$ binding energy.
Since the results depend on unknown details of $V_{eX^-}$ (due to the 
density-dependent polarization of the $X^-$ in the electric field of 
electrons), we make a (rough) approximation and instead of $V_{eX^-}$ 
use the pseudopotential of two distinguishable point charges with 
angular momenta $l_e$ and $l_{X^-}$.
The obtained spectra are quite close to the original ones and all 
contain the low lying bands as predicted by the CF model.
A much better fit is obtained for $V_{eX^-}$ including ($N$-dependent)
polarization effects.

It is apparent that only two types of states exhaust the entire low 
energy spectra shown in Fig.~\ref{fig03}: the MP states containing a 
decoupled $l_X=0$ exciton and the non-MP states containing an $X^-$.
None of the low energy states can be understood in terms of an excited 
($l_X\ne0$) exciton interacting with the excess $N-1$ electrons.
In particular, the bands of lowest energy states at $L=1$, 2, \dots, 
$N-3$ in Fig.~\ref{fig03}(acd) do not describe dispersion of a so-called 
``dressed exciton'' $X^*$ (charge neutral exciton with an enhanced mass 
due to the coupling to QE--QH pair excitations of the Laughlin 
$\nu={1\over3}$ fluid of $N-1$ excess electrons) as first suggested 
by Apalkov and Rashba\cite{apalkov} and reviewed in subsequent papers.
It is much more informative to interpret these $e$--$h$ states in 
terms of a well defined $X^-$ particle (with specified ${\cal Q}=-e$, 
$l=S-1$ or ${\cal L}=-1$, $\Delta$ as plotted in Fig.~\ref{fig02}, 
and $\tau^{-1}=0$) interacting with excess electrons through the well 
defined\cite{x-fqhe,x-cf} pseudopotential $V_{eX^-}$ yielding well 
defined Laughlin--Jastrow $e$--$X^-$ correlations and Laughlin 
quasiparticle excitations of a two-component incompressible 
``reference'' state, than to say that $k\ne0$ exciton is coupled 
in an undefined way to the Laughlin quasiparticles of an electron 
$\nu={1\over3}$ state.
The ``dressed exciton'' picture is simply wrong in describing the 
nature of the TQP of the system.
For example, the $X^*$ has zero charge and continuous energy 
spectrum instead of ${\cal Q}=-e$ and Landau quantized orbits 
of an $X^-$.
The reason why the suggestive idea of an $X^*$ does not work is 
that the coupling of a $k\ne0$ exciton (which has a nonzero in-plane 
electric dipole moment $\mu\propto k$) to electrons is too strong to 
be treated perturbatively.

\subsection{Small Layer Separation}
\label{secIV-2}

The knowledge of the nature of the TQP's of any system is essential 
for understanding its response to an external perturbation.
Since an $X^*$ is expected to behave differently than an $X^-$ when 
electron and hole layers are separated, the incorrect assumption of 
the ``dressed exciton'' picture at $d=0$ must result in incorrect 
interpretation of the $e$--$h$ states at $d>0$ as well.

At a small layer separation $d<\lambda$, all bound $e$--$h$ states
acquire a small electric dipole moment $\mu$, which is proportional 
to $d$ and oriented perpendicular the electron and hole planes.
These dipole moments result in a repulsive dipole--dipole interaction 
between $e$--$h$ complexes, which is proportional to $d^2/r^3$ at 
distance $r\gg d$.
While the electron--dipole $e$--$X$ repulsion is the reason for the 
decrease of the binding energy of an isolated $X^-$ at $0<d\ll\lambda$,
it can slightly extend the stability range of an $X^-$ embedded in 
a 2DEG (compared to Fig.~\ref{fig02}).

\begin{figure}[t]
\epsfxsize=3.40in
\epsffile{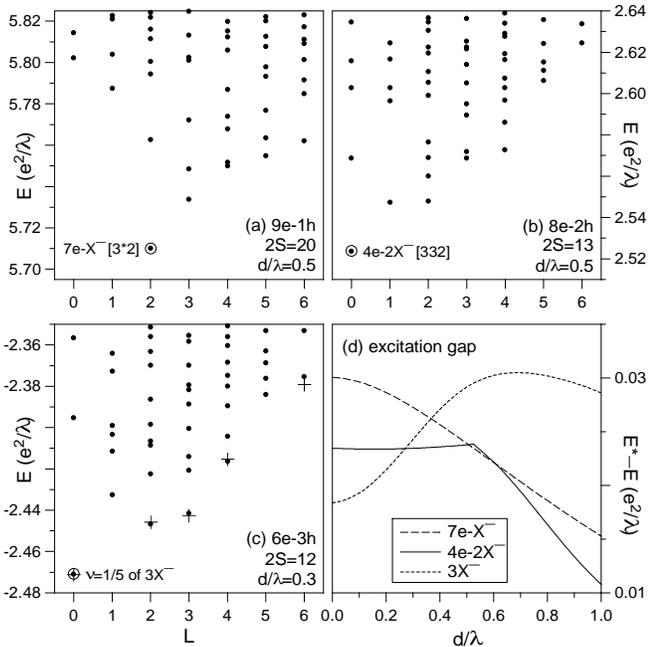}
\caption{
   (abc) The energy spectra (energy $E$ as a function of angular 
   momentum $L$) of electron--hole systems with Laughlin $e$--$X^-$ 
   correlations:
   (a) $7e$--$X^-$ ground state with [3*2] correlations in a $9e$--$h$
   system at the layer separation $d=0.5\lambda$;
   (b) $4e$--$2X^-$ incompressible ground state [332] in a $8e$--$2h$
   system at $d=0.5\lambda$;
   (c) Laughlin $\nu={1\over5}$ ground state of three $X^-$'s in a 
   $6e$--$3h$ system at $d=0.3\lambda$ (pluses show approximate $3X^-$
   energies).
   (d) The excitation gaps of ground states in frames (abc) as 
   a function of $d$.
   $\lambda$ is the magnetic length.}
\label{fig04}
\end{figure}

In the range of $d$ values for which the $X^-$ is bound, the $X^-$ 
dipole moment increases its total repulsion with electrons and other 
$X^-$'s.
It is possible that this increased repulsion could enhance the 
excitation gap of an incompressible fluid $e$--$X^-$ state.
Examples of different behavior of the gap are shown in Fig.~\ref{fig04}.
In Fig.~\ref{fig04}(a), the $9e$--$h$ ground state at $d=0.5\lambda$ 
is the $7e$--$X^-$ state with [3*2] correlations ($m_{X^-X^-}$ is 
undefined for only one $X^-$).
In the generalized CF picture, this state contains one QE$_{X^-}$ with
$l_{\rm QE}=2$ and a filled shell of electron CF's.
In Fig.~\ref{fig04}(b), the $8e$--$2h$ ground state at $d=0.5\lambda$ 
is the $4e$--$2X^-$ incompressible state [332].
In Fig.~\ref{fig04}(c), the $6e$--$3h$ ground state at $d=0.3\lambda$ 
is the Laughlin $\nu={1\over5}$ state of three $X^-$'s (here, pluses 
mark approximate $3X^-$ energies obtained by diagonalizing a system 
of three fermions each with energy $E_{X^-}$ and interacting through 
$V_{X^-X^-}$).
As shown in Fig.~\ref{fig04}(d), the excitation gaps of these three 
different Laughlin-correlated ground states behave differently as 
a function of $d$.
In particular, the gap of the $\nu={1\over5}$ state of $X^-$'s increases 
significantly up to $d=0.7\lambda$.

\section{Electron--Hole States at Large Layer Separation:
         Hole Weakly Coupled to Electron Fluid}
\label{secV}
It was shown by Chen and Quinn\cite{chen} that the opposite limit 
of $d\gg\lambda$ is easier to understand than that of $d<\lambda$, 
because of the vanishing $e$--$h$ interaction.
In this limit, the low lying states of the combined system are products 
of the Laughlin-correlated 2DEG and the decoupled hole.
The allowed angular momenta $L$ of the lowest energy band of the 
combined $e$--$h$ system result from addition of the angular momenta 
of the lowest energy electron states (containing a number of Laughlin 
quasiparticles) $L_e$ to the hole angular momentum $l_h=S$.

A decrease of $d$ to a few magnetic lengths $\lambda$ does not yet 
result in exciton binding because the length scale $D$ probed by the 
potential of a distant hole exceeds the average $e$--$e$ separation 
in the 2DEG.
While the $e$--$e$ interactions alone still completely determine the 
(Laughlin) correlations of the 2DEG, the valence band hole can now 
correlate with the quasiparticle excitations of the 2DEG due to their 
much lower density (compared to the electron density).
The hole repels positively charged QH's but can bind one or more 
negatively charged QE's (depending on the relative strength of the 
$h$--QE and QE--QE interactions) to form fractionally charged 
excitons (FCX), or ``anyonic ions,'' $h$QE$_n$.\cite{chen}
When $d$ is so large that the number of Laughlin quasiparticles in 
the 2DEG is conserved by the weak $e$--$h$ interaction, a discontinuity 
in the behavior of the system as a function of the magnetic field (or 
electron density) will occur at Laughlin filling factors $(2p+1)^{-1}$, 
because different type of TQP's can form depending on whether QE's are 
or are not present in the 2DEG.
The transition should be visible in PL, as the recombination of a free 
hole at $\nu<(2p+1)^{-1}$ can be distinguished from that of a hole bound 
into an $h$QE$_n$ complex at $\nu>(2p+1)^{-1}$.

\section{Electron--Hole States at Intermediate Layer Separation}
\label{secVI}
The TQP's of the $e$--$h$ system at a particular layer separation 
$d$ are by definition the most stable bound complexes (the ones with 
the largest binding energy) composed of smaller elementary particles 
or quasiparticles: a valence hole and either electrons or Laughlin 
excitations of the 2DEG.
To determine the most stable complexes at a particular value of $d$, 
the interactions between their sub-components must be studied.
Two-body interactions enter the many-body Hamiltonian through their 
pseudopotentials $V(L)$, defined as the pair interaction energy $V$ 
as a function of pair angular momentum $L$ (or another pair quantum 
number).\cite{haldane1,parentage}
The $e$--$e$, $e$--$h$, QE--QE, QH--QH, and QE--QH pseudopotentials 
are well-known\cite{haldane1,parentage,hierarchy} and (except for 
$e$--$h$) do not depend on $d$ for spatially separated electron and 
hole layers.
The simple form of single-particle wavefunctions in the lowest LL
results in very regular form of $V_{ee}(L)$ and $V_{eh}(L)$.
On a sphere, larger $L$ corresponds to smaller (larger) average 
separation of two charges of the same (opposite) sign, and thus
$V_{ee}$ increases and $|V_{eh}|$ decreases with increasing $L$.

\begin{figure}[t]
\epsfxsize=3.40in
\epsffile{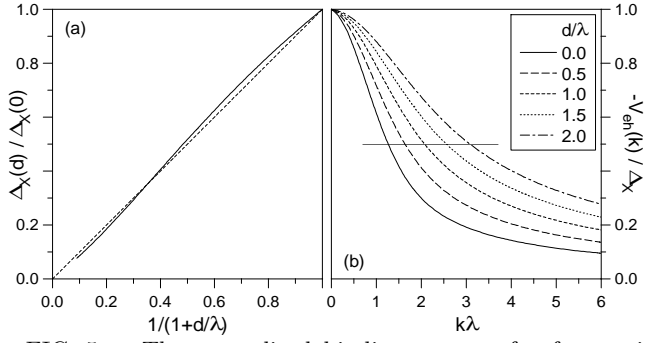}
\caption{
   The normalized binding energy of a free exciton, $\Delta_X(d)/
   \Delta_X(0)$, as a function of $(1+d/\lambda)^{-1}$ (a), and the 
   normalized electron--hole pseudopotentials $V_{eh}(k)/(-\Delta_X)$ 
   as a function of wavevector $k$ (b).
   $d$ is the separation between electron and hole layers, 
   and $\lambda$ is the magnetic length. 
}
\label{fig05}
\end{figure}

The dependence of $V_{eh}(L)$ on $d$ can be expressed in terms of 
the effective strength ($U$) and range ($D$) of the Coulomb potential 
of the hole (in its lowest-LL single-particle state) seen by an 
electron.
A measure of $U$ is the exciton binding energy $\Delta_X=V_{eh}(0)$.
As shown in Fig.~\ref{fig05}(a), $\Delta_X$ varies with $d$ roughly 
as $\Delta_X(d)=(1+d/\lambda)^{-1}\Delta_X(0)$, which means that the 
average $e$--$h$ separation in the exciton ground state is roughly 
$r_{eh}(d)=r_{eh}(0)+d$ rather than $\sqrt{r_{eh}^2(0)+d^2}$.
A measure of the range $D$ is an average $e$--$h$ distance $r_{eh}$
in the exciton state whose energy is half of the binding energy.
In Fig.~\ref{fig05}(b) we plot the normalized exciton pseudopotentials
as a function of wavevector $k=L/R$.
Since $r_{eh}$ is proportional\cite{gorkov} to $k$, and the value 
$k_{1/2}$ for which $V_{eh}(k_{1/2})=-{1\over2}\Delta_X$ in 
Fig.~\ref{fig05}(b) increases roughly linearly with $d$, we obtain 
the pair of relations, 
\begin{eqnarray}
   U&\propto&(1+d/\lambda)^{-1},
\nonumber\\
   D&\propto&d/\lambda,
\end{eqnarray}
describing the perturbing potentials $V_{UD}$ which can be achieved 
in bi-layer $e$--$h$ systems with different $d$.

Laughlin quasiparticles have more complicated charge density profiles 
than electrons or holes in the lowest LL.
This internal structure is reflected in the oscillations of the QE 
and QH pseudopotentials at the values of $L$ corresponding to small 
average separation between the QE or QH and the second particle.
For example, despite Laughlin quasiparticles being charge excitations, 
neither QE--QE nor QH--QH interaction is generally repulsive.
\cite{hierarchy,yi}
On the contrary, the QE$_2$ molecule (the state with maximum $L$, i.e. 
minimum average QE--QE separation) is either the ground state or a very 
weakly excited state of two QE's (the numerical results for finite systems 
are not conclusive).\cite{hierarchy}

\begin{figure}[t]
\epsfxsize=3.40in
\epsffile{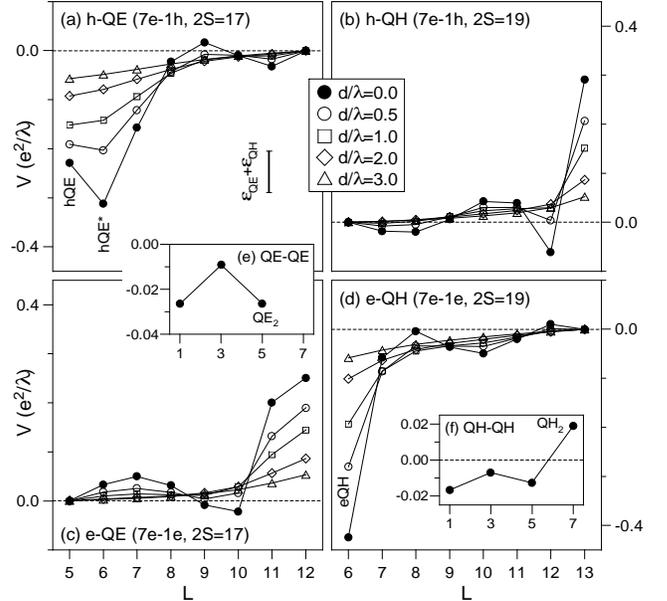}
\caption{
   The pseudopotentials (pair energy $V$ as a function of pair angular 
   momentum $L$) of the interaction between quasiparticles (quasielectron 
   QE and quasihole QH) of the seven-electron Laughlin $\nu={1\over3}$ 
   state and an additional charge (electron or hole) on a parallel layer 
   separated by $d$.
   The QE--QE and QH--QH pseudopotentials for $N=7$ are shown in inset 
   frames (ef).
   $\varepsilon_{\rm QE}$ and $\varepsilon_{\rm QH}$ are the QE and QH
   energies and $\lambda$ is the magnetic length. 
}
\label{fig06}
\end{figure}

In order to calculate the pseudopotentials $V_{h{\rm QE}}(L)$ and 
$V_{h{\rm QH}}(L)$ associated with the interaction between Laughlin 
quasiparticles (QE or QH) of a $\nu={1\over3}$ fluid and a hole moving 
in a parallel plane separated by an arbitrary distance $d$, we use the 
following procedure.
A finite $Ne$--$h$ system is diagonalized at the monopole strength 
$2S$ corresponding to a single QE or QH in the 2DEG (in the absence 
of the interaction with the hole).
To assure that the interaction between the hole and the 2DEG is weak 
compared to the energy $\varepsilon_{\rm QE}+\varepsilon_{\rm QH}$ 
($\approx0.1\,e^2/\lambda$ for an infinite system) needed to create 
additional QE--QH pairs in the 2DEG, the charge of the hole is set 
to $e/\epsilon$ where $\epsilon\gg1$.
This guarantees that the lowest band of $Ne$--$h$ states contain 
exactly one QE or QH interacting with the hole.
The pseudopotentials $V_{h{\rm QE}}(L)$ and $V_{h{\rm QH}}(L)$ 
are calculated by subtracting from the lowest eigenenergies the 
constant energy of the 2DEG and the energy of interaction between 
the hole and the uniform-density $\nu={1\over3}$ fluid, and multiplying 
the difference by $\epsilon$.
If $\epsilon$ is sufficiently large, the pseudopotentials calculated 
in this way (and shown in Fig.~\ref{fig06}(ab)) do not depend on 
$\epsilon$ and describe the interaction between the hole of full 
charge $+e$ and the Laughlin quasiparticle. 

\begin{figure}[t]
\epsfxsize=3.40in
\epsffile{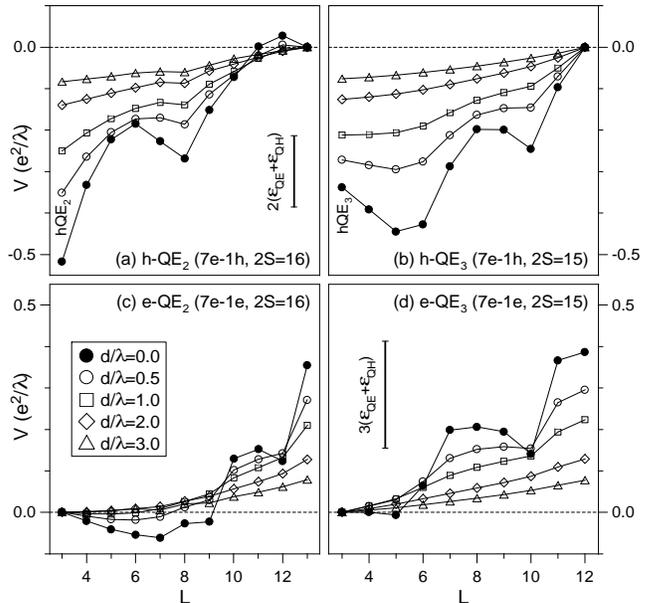}
\caption{
   The pseudopotentials (pair energy $V$ versus pair angular momentum 
   $L$) of the interaction between molecules consisting of two or
   three quasielectrons (QE$_2$ and QE$_3$) of the Laughlin 
   $\nu={1\over3}$ state and an additional charge (electron or hole) 
   on a parallel layer separated by $d$.
   $\varepsilon_{\rm QE}$ and $\varepsilon_{\rm QH}$ are the QE and QH
   energies and $\lambda$ is the magnetic length. 
}
\label{fig07}
\end{figure}

A similar procedure has been used to calculate the pseudopotentials 
$V_{e{\rm QH}}(L)$ and $V_{e{\rm QE}}(L)$ of the interaction 
between quasiparticles and an electron moving in a parallel layer
[Fig.~\ref{fig06}(cd)], and the pseudopotentials $V_{h{\rm QE}_n}(L)$ 
and $V_{e{\rm QE}_n}(L)$ involving the QE$_2$ and QE$_3$ molecules 
(Fig.~\ref{fig07}).
From such calculation, the binding energies and PL oscillator strengths 
of all $h$QE$_n$ FCX's are obtained to determine under what circumstances 
(layer separation, density, temperature, etc.) various FCX's can occur 
and contribute to the PL spectrum.

The pseudopotentials of a single QE and QH of a seven-electron fluid 
($N=7$) interacting with a hole or an electron on a parallel layer 
are shown in Fig.~\ref{fig06}(ab) for a number of different layer 
separations $d$.
The allowed pair angular momenta $L$ result from addition of individual
angular momenta of the quasiparticles,\cite{hierarchy} $l_{\rm QE}=
l_{\rm QH}=N/2$, and the particles in the second layer, $l_e=l_h=S$.
Since the length scale $D$ probed by the potential of the hole (electron) 
decreases when it is brought closer to the 2DEG, structure appears for 
$d<\lambda$ in all pseudopotentials (at $L$ corresponding to small 
average separation).
For example, the $h$--QE ground state for $d<\lambda$ occurs at 
$L>l_h-l_{\rm QE}$, i.e. not at the minimum allowed average $h$--QE 
separation.
Similarly as in QE--QE and QH--QH pseudopotentials\cite{hierarchy} 
(see also Fig.~\ref{fig06}(ef) for $N=7$), the oscillations of 
particle--quasiparticle pseudopotentials reflects structure in QE 
and QH charge density.

\begin{figure}[t]
\epsfxsize=3.40in
\epsffile{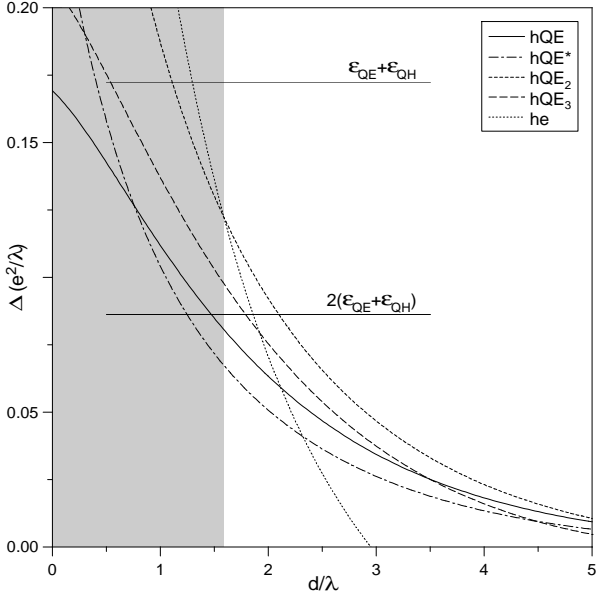}
\caption{
   The binding energy $\Delta$ of fractionally charged excitons 
   $h$QE$_n$ as a function of layer separation $d$, calculated 
   for the $8e$--$h$ system with a fixed number of Laughlin 
   quasiparticles in the $8e$ electron system ($\epsilon\gg1$; 
   see text).
   $\lambda$ is the magnetic length. 
   The $he$ state contains an exciton and originates from the 
   multiplicative state at $d=0$.
   In the shaded part of the graph, the $he$ has the largest binding 
   energy and the $h$QE$_n$ complexes do not form.}
\label{fig08}
\end{figure}

All pseudopotentials in Figs.~\ref{fig06} and \ref{fig07} have been 
arbitrarily shifted in energy so that they vanish for the pair state 
of the largest average separation.
The more accurate estimate of the $h$--QE pseudopotential parameters
at the two smallest values of $L$, i.e. the binding energy $\Delta$ of 
the $h$QE and $h$QE* complexes with the smallest and the next smallest 
average $h$--QE separation (the $h$QE* complex is important in discussion 
\cite{bilayer-pl} of PL) gives the curves plotted in Fig.~\ref{fig08}.
The interaction of the 2DEG at $\nu\approx{1\over3}$ with an additional 
charge (hole or electron) can be considered weak only at about 
$d>1.5\lambda$.
In this regime, the 2DEG responds to the perturbation introduced by 
a distant charge by screening it with already existing Laughlin 
quasiparticles to form bound FCX's, $h$QE or $e$QH.
A discontinuity occurs at $\nu={1\over3}$, because the QE's that can 
be bound to a hole exist only at $\nu>{1\over3}$, and the QH's that 
can be bound to an electron occur only at $\nu<{1\over3}$.
Fig.~\ref{fig08} shows that at $d<1.5\lambda$ the energy of $h$--QE 
(and $e$--QH) attraction exceeds $\varepsilon_{\rm QE}+\varepsilon_{\rm 
QH}$ and the QE--QH pairs are spontaneously created in the 2DEG to screen 
the hole (or electron) charge at any value of $\nu\approx{1\over3}$.

Whether only one QE--QH pair will be spontaneously created to form 
$h$QE, or if additional QE--QH pairs will be created to form larger 
FCX's (e.g., $h$QE$\,\rightarrow h$QE$_2+$QH) depends on $V_{h{\rm 
QE}_2}$ and $V_{h{\rm QE}_3}$.
Since $V_{\rm QE-QE}$ has a minimum at $L=2l_{\rm QE}-1$ (${\cal R}=1$)
and a maximum at $L=2l_{\rm QE}-3$ (${\cal R}=3$), two or three QE's 
can form QE$_2$ or QE$_3$ molecules.
Even if the QE$_2$ and QE$_3$ molecules do not turn out to be the 
absolute two- or three-QE ground states in the absence of an additional 
attractive potential, they both will be metastable due to the energy 
barrier at ${\cal R}=3$, i.e. a finite energy gap to separate two QE's.
Both QE$_2$ and QE$_3$ can bind to a hole, and (because of the barrier
in $V_{\rm QE-QE}$) the resulting FCX's, $h$QE$_2$ and $h$QE$_3$, could 
be expected to be quite stable even at $d\gg\lambda$.

The pseudopotentials describing interaction of the QE$_2$ and QE$_3$ 
molecules with a hole and an electron are shown in Fig.~\ref{fig07}.
Somewhat unexpectedly, they show that QE$_2$ is more strongly attracted 
to the hole than QE$_3$, which suggests that the $h$QE$_3$ is not stable 
($h$QE$_3\rightarrow h$QE$_2+$QE). 
Since the $h$--QE$_2$ attraction is also stronger than $h$--QE in 
Fig.~\ref{fig06}, both $h$QE and $h$QE$_2$ are stable FCX's.

The binding energies $\Delta$ of all $h$QE$_n$ complexes calculated 
in the $8e$--$h$ system are plotted as a function of $d$ in 
Fig.~\ref{fig08}.
The binding energy $\Delta$ of an $h$QE$_n$ state is defined as the 
energy of attraction between the hole and $n$ QE's.
For the excitonic state $he$ (in which a hole binds a whole ``real'' 
electron to form an $e$--$h$ pair weakly coupled to the remaining $N-1$ 
electrons at at $2S=3(N-2)$, i.e. at $\nu={1\over3}$) with energy 
$E_{he}$, $\Delta_{he}$ is defined as a difference between $E_{he}$ 
and the state in which the hole is completely decoupled from all $N$ 
electrons (which at $2S=3(N-2)$ form a state with three Laughlin QE's).
Note that $\Delta_{he}$ is not equivalent to the binding energy of a 
free exciton (it is not equal to the $e$--$h$ attraction but also 
includes the energy needed to remove an electron from the Laughlin 
state so that it can be bound to the hole).

The $h$QE$_2$ is the most strongly bound FCX in entire range of $d$
(at least up to $d=10\lambda$), and hence it is expected to form in 
the presence of excess QE's at $\nu>{1\over3}$.
It can be seen in Fig.~\ref{fig08} that $\Delta_{h{\rm QE}_2}>
\varepsilon_{\rm QE}+\varepsilon_{\rm QH}$ at $d<\lambda$, and two 
QE--QH pairs are spontaneously created in the 2DEG to form $h$QE$_2$ 
even at $\nu<{1\over3}$.
However, at such small $d$, neutral ($X$) and charged excitons ($X^-$) 
composed of a hole and one or two ``real'' electrons of charge $-e$ 
(rather than Laughlin QE's of charge $-{1\over3}e$) are more stable 
complexes than $h$QE$_2$. 
The transition from fractional to ``normal'' exciton phase occurs at
$d\approx1.5\lambda$, that is at the the crossing of $\Delta_{h{\rm 
QE}_2}$ and $\Delta_{he}$ in Fig.~\ref{fig08} (the shaded rectangle 
marks the ``normal'' exciton phase).

\section{Numerical Energy Spectra}
\label{secVII}
\begin{figure}[t]
\epsfxsize=3.40in
\epsffile{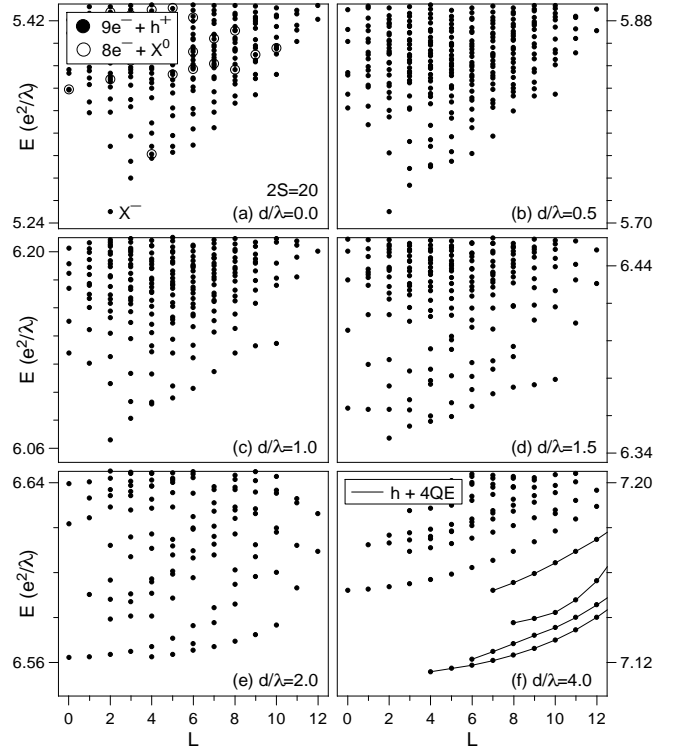}
\caption{
   The energy spectra (energy $E$ vs.\ angular momentum $L$) of 
   the $9e$--$h$ system calculated on a Haldane sphere with monopole 
   strength $2S=20$ for different layer separations $d$.
   The open circles mark the multiplicative states at $d=0$.
   $\lambda$ is the magnetic length. 
}
\label{fig09}
\end{figure}

\begin{figure}[t]
\epsfxsize=3.40in
\epsffile{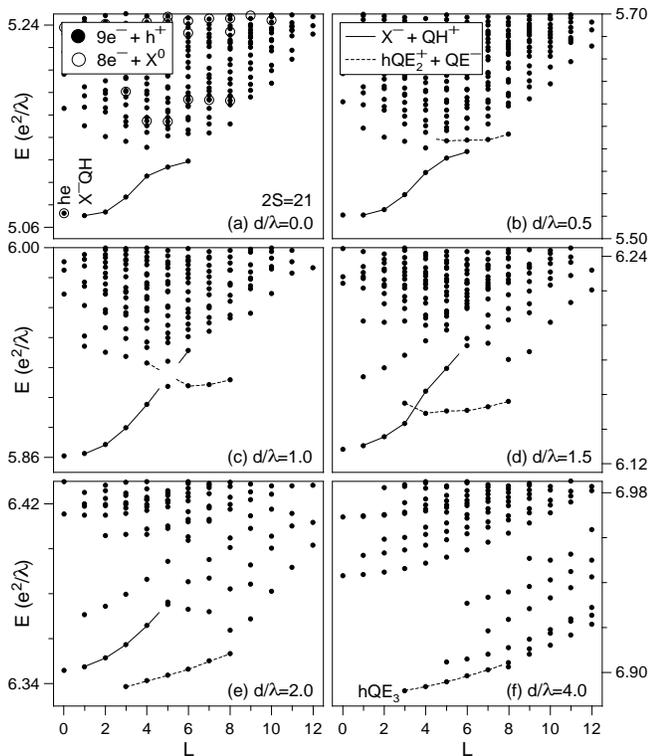}
\caption{
   The energy spectra (energy $E$ vs.\ angular momentum $L$) of 
   the $9e$--$h$ system calculated on a Haldane sphere with monopole 
   strength $2S=21$ for different layer separations $d$.
   The open circles mark the multiplicative states at $d=0$.
   $\lambda$ is the magnetic length. 
}
\label{fig10}
\end{figure}

\begin{figure}[t]
\epsfxsize=3.40in
\epsffile{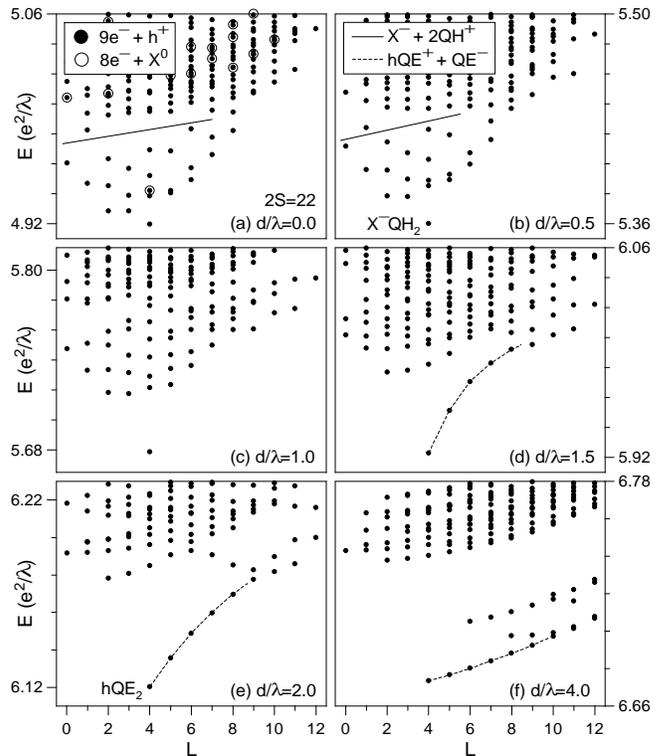}
\caption{
   The energy spectra (energy $E$ vs.\ angular momentum $L$) of 
   the $9e$--$h$ system calculated on a Haldane sphere with monopole 
   strength $2S=22$ for different layer separations $d$.
   The open circles mark the multiplicative states at $d=0$.
   $\lambda$ is the magnetic length. 
}
\label{fig11}
\end{figure}

\begin{figure}[t]
\epsfxsize=3.40in
\epsffile{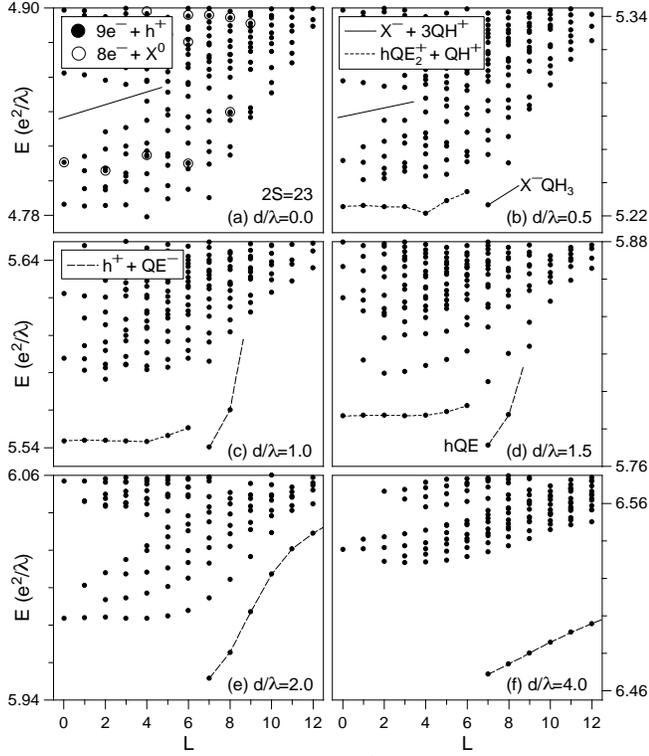}
\caption{
   The energy spectra (energy $E$ vs.\ angular momentum $L$) of 
   the $9e$--$h$ system calculated on a Haldane sphere with monopole 
   strength $2S=23$ for different layer separations $d$.
   The open circles mark the multiplicative states at $d=0$.
   $\lambda$ is the magnetic length. 
}
\label{fig12}
\end{figure}

\begin{figure}[t]
\epsfxsize=3.40in
\epsffile{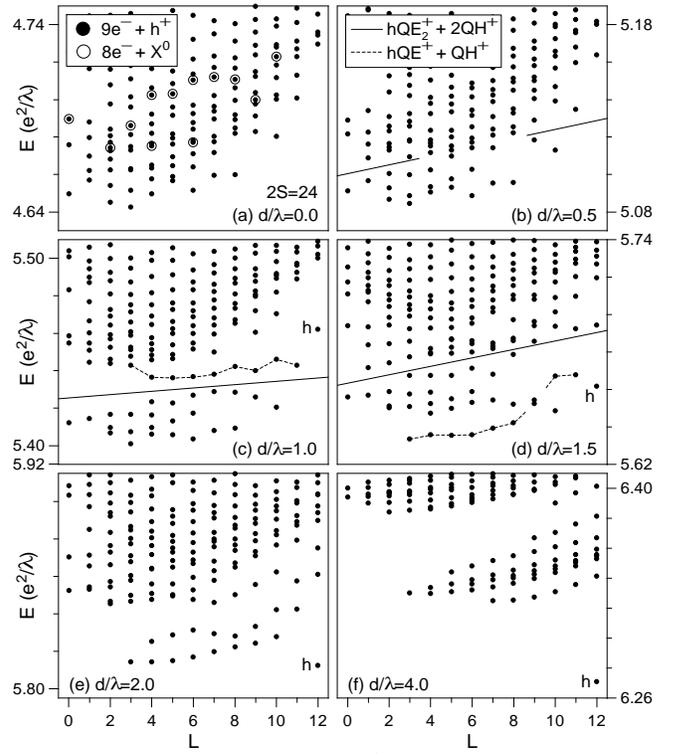}
\caption{
   The energy spectra (energy $E$ vs.\ angular momentum $L$) of 
   the $9e$--$h$ system calculated on a Haldane sphere with monopole 
   strength $2S=24$ for different layer separations $d$.
   The open circles mark the multiplicative states at $d=0$.
   $\lambda$ is the magnetic length. 
}
\label{fig13}
\end{figure}

Using a modified Lanczos algorithm, we have been able to quickly 
and exactly diagonalize Hamiltonians of dimensions up to $\sim10^6$.
This allowed calculation of energy and PL spectra of $Ne$--$h$ systems 
with $N\le9$ and at the values of $2S$ up to $3(N-1)$, corresponding 
to the hole interacting with the Laughlin $\nu={1\over3}$ state of $N$ 
electrons.
The $9e$--$h$ energy spectra (energy $E$ as a function of angular 
momentum $L$) obtained for different values of $2S$ and $d$ have been 
shown in Figs.~\ref{fig09}--\ref{fig13}.
In all figures, the open circles at $d=0$ mark the MP states, in which 
a $k=0$ exciton is decoupled from remaining electrons.

\paragraph*{\rm $2S=20$ (Fig.~\ref{fig09}):}
The spectrum for $d=0.5\lambda$ has already been shown in 
Fig.~\ref{fig04}(a).
At a low layer separation ($d\le\lambda$), the ground state at $L=2$ is 
the $7e$--$X^-$ Laughlin-correlated state [3*2].
In the CF picture of this state, seven electrons fill completely their 
CF shell of $l_e^*=S-(N-2)=3$ and the $X^-$ becomes a ``quasielectron'' 
(QE$_{X^-}$) with $l_{X^-}^*=l_e^*-1=2$.

At $d\approx2\lambda$, the $X^-$ unbinds and the $7e$--$X^-$ fluid 
undergoes reconstruction.
Since four QE's of the nine electron Laughlin $\nu={1\over3}$ state 
cannot all bind to the hole, the ground state at the intermediate 
separations ($1.5\lambda\le d\le4\lambda$) does not correspond to 
a single bound FCX.
Instead, the low lying states describe different unbound states of 
the hole and four QE's which occur in this particular finite-size 
system (but have no significance in an infinite system).

At very large separations ($d>4\lambda$) the spectrum simplifies due
to the weakening of $h$--QE interactions, and well defined bands develop 
in the low energy spectrum.
In these bands, the hole with angular momentum $l_h=S=10$ is weakly 
coupled to different states of four QE's in Laughlin fluid of nine 
electrons.
Each QE has $l_{\rm QE}=S-(N-1)+1=3$, the allowed total angular momenta 
of four QE's are $L_{4{\rm QE}}=0$, 2, 3, 4, and 6, and the $h$--$4$QE 
bands start at $L=|l_h-L_{4{\rm QE}}|=10$, 8, 7, 6, and 4.
All these bands (except for the band with $L_{4{\rm QE}}=0$ and $L=10$ 
which has too high energy at $d\le4\lambda$) are marked in 
Fig.~\ref{fig09}(e) with solid lines.
The states within each $h$--$4$QE band become degenerate at 
$d/\lambda\rightarrow\infty$.

\paragraph*{\rm $2S=21$ (Fig.~\ref{fig10}):}
The spectrum for $d=0$ has already been shown in Fig.~\ref{fig03}(d).
As explained in Sec.~\ref{secIV-1}, the lowest energy state at $L=0$
is the MP state, and the band of states connected with solid lines 
describe excitonic spectrum of an $X^-$ interacting with one QH of 
the two-component Laughlin [3*2] state of the $7e$--$X^-$ fluid.
Clearly, this band survives also in the spectra at $d>0$.
The $X^-$--QH states at the largest $E$ and $L$ (in which the $X^-$ 
is far away from the QH) gain energy and fall into the continuum at 
$d\approx\lambda$, i.e. when the $X^-$ is expected to unbind.
In the $X^-$QH state at $L=1$, the positively charged QH is closely 
bound to the $X^-$ which reduces its total charge (the total charge 
of the $X^-$QH is ${\cal Q}=-{2\over3}$, compared to ${\cal Q}=-1$ 
for a free $X^-$), and stabilizes the $X^-$QH state up to at least 
$d=2\lambda$.

\begin{figure}[t]
\epsfxsize=3.40in
\epsffile{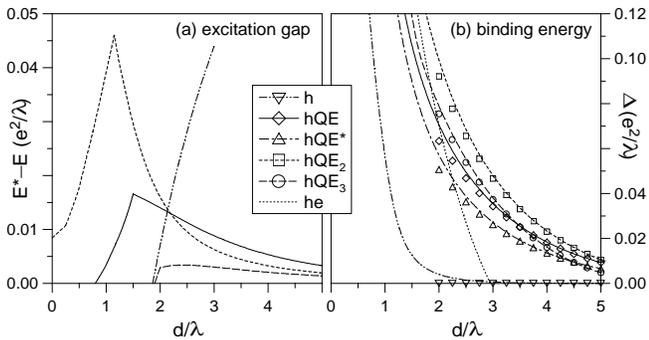}
\caption{
   The excitation gap $E^*-E$ (a), and binding energy $\Delta$ (b) 
   of fractionally charged excitons $h$QE$_n$ as a function of layer 
   separation $d$, calculated for the $8e$--$h$ system.
   $E_X$ is the exciton energy and $\lambda$ is the magnetic length. 
   The $he$ state contains an exciton and originates from the 
   multiplicative state at $d=0$.}
\label{fig14}
\end{figure}

The band of states marked with dashed lines contains the $h$QE$_2$
(most stable of all FCX's) interacting with the third QE (note that 
here QE denotes quasielectron of the Laughlin $\nu={1\over3}$ state 
of nine electrons, not of the two-component $7e$--$X^-$ state [3*2]).
The allowed angular momenta of the $h$QE$_2$--QE pair can be obtained
by adding two $l_{\rm QE}\equiv S-(N-1)+1={7\over2}$ to obtain 
$l_{{\rm QE}_2}\equiv2l_{\rm QE}-1=6$ and then adding to it $l_h={21
\over2}$ to obtain $l_{h{\rm QE}_2}\equiv|l_h-l_{\rm QE}|={9\over2}$. 
Finally, adding $l_{h{\rm QE}_2}$ to $l_{\rm QE}$ as if they were
completely distinguishable particles gives $|l_{h{\rm QE}_2}-l_{\rm 
QE}|\le L\le l_{h{\rm QE}_2}+l_{\rm QE}$, or $1\le L\le8$.
However, because $h$QE$_2$ contains two QE's which are fermions 
indistinguishable from the third QE, the exclusion principle 
forbids the states with $L=1$ and 2 ($L_{3{\rm QE}}\le{15\over2}$ 
so that adding it to $l_h={21\over2}$ cannot give $L<3$).
The resulting band at $L=3$, 4, \dots, 8 indeed appears in 
the $9e$--$h$ spectrum at $\lambda\le d\le2\lambda$.
The dependence of $E$ on $L$ within this band is (up to a 
constant shift in energy) the $h$QE$_2$--QE pseudopotential, 
$V_{h{\rm QE}_2-{\rm QE}}(L)$.
Since $h$QE$_2$ and QE have opposite charge (opposite angular momentum),
the decrease of $V_{h{\rm QE}_2-{\rm QE}}$ as a function of $L$ at
$d\le1.5\lambda$ signals the $h$QE$_2$--QE repulsion, consistent with
the result in Fig.~\ref{fig08} that $h$QE$_3$ does not bind.
At $d\ge4\lambda$, the low energy bands contain low energy three-QE 
eigenstates weakly coupled (bound) to the hole.
At $d=4\lambda$, the lowest of those band contains the QE$_3$ molecule,
with $l_{{\rm QE}_3}\equiv3l_{\rm QE}-3={15\over2}$, and the ground 
state is $h$QE$_3$ at $l_{h{\rm QE}_3}\equiv|l_h-l_{{\rm QE}_3}|=3$.
At even much larger $d$, the ground state consist of the three-QE ground 
state (which for $N=9$ at $2S=21$ occurs at $L_{3{\rm QE}}={5\over2}$) 
virtually decoupled from the hole.
Note that the fact that $h$QE$_3$ is the $9e$--$h$ ground state in 
Fig.~\ref{fig10}(ef) does not mean that it is a stable bound complex 
in an infinite system.
The comparison of binding energies in Figs.~\ref{fig08} and \ref{fig14}(b) 
shows that if only the third QE could move away from the $h$QE$_2$ (which 
it cannot do in a finite system in Fig.~\ref{fig10}), the $h$QE$_3$ would 
unbind at any $d$.

\paragraph*{\rm $2S=22$ (Fig.~\ref{fig11}):}
At small $d$, the low energy states contain an $X^-$ interacting with 
two QH's of the Laughlin [3*2] state of the $7e$--$X^-$ fluid.
The angular momenta of an electron ''quasihole'' QH$_e$ (that we will 
denote here simply by QH) and an $X^-$ ``quasielectron'' QE$_{X^-}$ 
(denoted simply by $X^-$) are obtained from the generalized CF picture: 
$l_e^*\equiv S-(N-2)=4$ and $l_{X^-}^*\equiv l_e^*-1=3$.
The two-QH states can have $L_{2{\rm QH}}=2l_e^*-{\cal R}=1$, 3, 5, 
or 7.
Adding allowed $L_{2{\rm QH}}$ to $l_{X^-}^*$ gives allowed total 
$X^-$--2QH angular momenta $L=0$, 1, $2^2$, $3^2$, \dots, 9, and 10.
Indeed, these multiplets form the lowest energy band of states at 
$d\le0.5\lambda$ separated from higher states by solid lines in 
Fig.~\ref{fig11}(ab). 

The lowest state in this band (the $9e$--$h$ ground state) is the 
bound $X^-$QH$_2$ state, at angular momentum $l_{X^-{\rm QH}_2}\equiv
|l_{{\rm QH}_2}-l_{X^-}^*|=4$.
The lowest MP state at $d=0$ and $L=4$ (marked with an open circle) 
has higher energy than $X^-$QH$_2$. 
It contains a $k=0$ exciton decoupled from one Laughlin QH of the 
eight electron system.

At $d>\lambda$, the low energy band of states develops at $L\ge4$.
These states contain an $h$QE interacting with the second QE
(this interaction is attractive, because $V_{h{\rm QE}-{\rm QE}}$ 
increases as a function of $L$, and $h$QE and QE have opposite charge).
The lowest state is the bound $h$QE$_2$ state, whose angular momentum 
$l_{h{\rm QE}_2}=4$ results from addition of two $l_{\rm QE}=4$ to 
obtain $l_{{\rm QE}_2}=7$, and then adding to it $l_h=11$.
Note that because $h$QE$_2$ has the same angular momentum $L={1\over2}
(N-1)$ as $X^-$QH$_2$, the transition from one state to the other is
continuous.
It is apparent from the dependence of PL intensity\cite{bilayer-pl} 
on $d$ that it occurs about $d\approx1.66\lambda$.

\paragraph*{\rm $2S=23$ (Fig.~\ref{fig12}):}
At small $d$, the low energy states contain an $X^-$ interacting with 
three QH's of the Laughlin [3*2] state of the $7e$--$X^-$ fluid.
The generalized CF picture uses $l_e^*\equiv S-(N-2)={9\over2}$ and 
$l_{X^-}^*\equiv l_e^*-1={7\over2}$, and predicts $L=1$, $2^4$, $3^6$, 
\dots, 13 for this band.
Indeed, at least at small $L$, these $X^-$--3QH states can be 
identified in Fig.~\ref{fig12}(ab).
The angular momentum of a bound $X^-$QH$_3$ results from adding three 
$l_{{\rm QH}_3}=3l_e^*-3$ to $l_{X^-}^*$ to obtain $l_{h{\rm QH}_3}=
|l_{{\rm QH}_3}-l_{X^-}^*|=7$.
Although most likely $X^-$QH$_3$ is the lowest state at $L=7$ in 
Fig.~\ref{fig12}(a), it has higher energy than other states and thus it 
is unstable (due to the short range of QH--QH repulsion;\cite{hierarchy} 
see also Fig.~\ref{fig06}(f) for the QH--QH pseudopotential in a 
seven-electron system).

At $d>\lambda$, the $X^-$ unbinds and the $X^-$--3QH band undergoes
reconstruction. 
At $d\approx\lambda$, two competing low energy bands occur in the spectra 
in Fig.~\ref{fig12}(bcd).
One describes the hole with $l_h\equiv S={23\over2}$ and the QE 
with $l_{\rm QE}\equiv S-(N-1)+1={9\over2}$ interacting through 
a pseudopotential similar to that in Fig.~\ref{fig06}(a).
This band has $L\ge|l_h-l_{\rm QE}|=7$, and the lowest two states 
(at $L=7$ and 8) are $h$QE and $h$QE*.
The second band involves an additional QE--QH pair and describes the 
$h$QE$_2$ with $l_{h{\rm QE}_2}\equiv|l_h-l_{{\rm QE}_2}|=|l_h-
(2l_{\rm QE}-1)|={7\over2}$ interacting with the QH with $l_{\rm QH}
\equiv S-(N-1)={7\over2}$.
The angular momenta $L$ obtained by adding $l_{h{\rm QE}_2}$ and 
$l_{\rm QH}$ satisfy $|l_{h{\rm QE}_2}-l_{\rm QH}|\le L\le l_{h{\rm 
QE}_2}+l_{\rm QH}$, i.e. $0\le L\le7$.
Because of the ``hard core'' of $V_{{\rm QE}-{\rm QH}}$ (the QE--QH 
state at $L=1$ does not occur\cite{parentage}), the $h$QE$_2$--QH 
state at the highest value of $L$ is forbidden, and the $h$QE$_2$--QH 
band has $L=0$, 1, 2, \dots, 6.
We showed in Sec.~\ref{secVI} that creation of an additional QE--QH 
pair to bind the second QE to $h$QE and form $h$QE$_2$ is energetically 
favorable at $d\le\lambda$ (see the crossing of $\Delta_{h{\rm QE}_2}$
and $2(\varepsilon_{\rm QE}+\varepsilon_{\rm QH})$ in Fig.~\ref{fig08}).
Indeed, in Fig.~\ref{fig12} the $h$QE state crosses the $h$QE$_2$--QH 
band and becomes the $9e$--$h$ ground state at $d\approx\lambda$.

\paragraph*{\rm $2S=24$ (Fig.~\ref{fig13}):}
At small $d$, the lowest states contain an $X^-$ interacting with 
four QH's of the [3*2] state of the $7e$--$X^-$ fluid.
This band is fairly broad and overlaps with higher ones, containing 
additional QE--QH pairs.

In the $L\equiv S=12$ ground state at very large $d$, the hole is 
decoupled from the $\nu={1\over3}$ state of $N=9$ electrons. 
The excitation gap in Fig.~\ref{fig13}(f) is the Laughlin gap of the 
2DEG, and the first excited band describes the interaction of the 
hole with an additional QE--QH pair [$l_{\rm QE}\equiv S-(N-1)+1=5$ 
and $l_{\rm QH}\equiv S-(N-1)=4$], and it contains $L=3$, $4^2$,
$5^3$, $6^4$, $7^5$, \dots, resulting from adding $2\le L_{{\rm QE}
-{\rm QH}}\le9$ to $l_h=12$.
At $d\approx1.5\lambda$, the $h$--QE attraction exceeds both the 
magneto-roton energy (QE--QH attraction) and the $\varepsilon_{\rm QE}+
\varepsilon_{\rm QH}$ gap, and the lowest energy band at $3\le L\le11$
in Fig.~\ref{fig13}(d) contains a bound $h$QE state with $l_{h{\rm QE}}
\equiv|l_h-l_{\rm QE}|=7$ interacting with a QH.
At $d\approx\lambda$, the $h$--QE attraction becomes even stronger and
the $h$QE$_2$ state with $l_{h{\rm QE}_2}\equiv|l_h-l_{{\rm QE}_2}|=3$
is formed.
The lowest band in Fig.~\ref{fig13}(c) describes the interaction of 
an $h$QE$_2$ state with two QH's.
The addition of $l_{h{\rm QE}_2}$ and two $l_{\rm QH}$'s gives a band 
at $L=0$, 1, $2^3$, \dots, 10 observed in Fig.~\ref{fig13}(c).

In Fig.~\ref{fig14} we present the data regarding the stability of 
different FCX's, extracted from the $8e$--$h$ spectra, similar 
those in Figs.~\ref{fig09}--\ref{fig13}.
We have checked that the curves plotted here for $N=8$ are very close 
to those obtained for $N=7$ or 9, so that all important properties of 
an extended system can be understood from a rather simple $8e$--$h$
computation.
In two frames, for each $h$QE$_n$ we plot: 
(a) the excitation gap $E^*-E$ above the $h$QE$_n$ ground state, and
(b) the binding energy $\Delta$.
The excitation gaps are obtained from the spectra at $2S=3(N-1)-n$ in 
which isolated $h$QE$_n$ complexes occur.
The binding energy $\Delta$ is defined in such way that $E_{h{\rm QE}_n}
=E_{{\rm QE}_n}+V_{h-{\rm LS}}-\Delta$, where $E_{h{\rm QE}_n}$ is the 
energy of the $Ne$--$h$ system in state $h$QE$_n$ calculated at 
$2S=3(N-1)-n$, $E_{{\rm QE}_n}$ is the energy of the $Ne$ system in 
state QE$_n$ calculated at the same $2S=3(N-1)-n$, and $V_{h-{\rm LS}}$ 
is the self-energy of the hole in Laughlin $\nu={1\over3}$ ground state 
at $2S=3(N-1)$.
As described in Sec.~\ref{secVI}, $V_{h-{\rm LS}}$ is calculated by 
setting the hole charge to a very small fraction of $+e$ so that it 
does not perturb the Laughlin ground state.

The lines in Fig.~\ref{fig14} show data obtained from the spectra 
similar to those in Figs.~\ref{fig09}--\ref{fig13}, i.e. including all 
effects of $e$--$h$ interactions.
For comparison, with symbols we have shown the data plotted previously 
in Fig.~\ref{fig08}, where very small hole charge $e/\epsilon$ was 
used in the calculation to assure that, at any $d$, the obtained low 
energy eigenstates are given exactly by the $h$QE$_n$ wavefunction.
At $d>\lambda$, very good agreement between binding energies calculated 
for $\epsilon=1$ (lines) and $\epsilon\gg1$ (symbols) confirms our 
identification of $h$QE$_n$ states in low energy $Ne$--$h$ spectra.
At $d<\lambda$ the two calculations give quite different results 
which confirms that the description of actual $Ne$--$h$ eigenstates
in terms of the hole interacting with Laughlin quasiparticles of the
2DEG is inappropriate (the correct picture is that of a two-component 
$e$--$X^-$ fluid).

The formation of $h$QE$_n$ complexes at $d$ larger than about $1.5
\lambda$ can be seen most clearly in the dependence\cite{bilayer-pl} 
of their PL intensity on $d$.
Although $d$ is the only tunable parameter in an $e$--$h$ system,
the transition from ``integrally'' to ``fractionally'' charged 
exciton phase occurs in the phase space of two parameters, $D$ 
and $U$, which define the perturbation potential $V_{UD}$.
Different combinations of $U$ and $D$ are possible in systems where 
the hole is replaced by an electrode (STM) or a charged impurity.
\cite{rezayi1,fox}
The relation between $U$ and $D$ in realistic $e$--$h$ systems depends
somewhat on the magnetic field and electron density (because of the 
asymmetric inter-LL scattering for electrons and holes), and/or on the 
widths of electron and hole layers.
We have calculated similar dependences to those in Fig.~\ref{fig14} 
for the $e$--$h$ interaction multiplied by a constant, $\epsilon^{-1}
V_{eh}$, and found that the phase transition occurs in every case.
The critical layer separation depends on $\epsilon$ and equals
 $d/\lambda=0.84$, 1.66, 2.25, 2.61, and 2.95, for $\epsilon^{-1}=0.5$, 
1, 1.5, 2, and 2.5, respectively.

The analysis of the characteristics of $h$QE$_n$ complexes plotted in 
Fig.~\ref{fig14} (and the good agreement of the actual binding energies 
with those obtained for $\epsilon\gg1$) confirms that the most important 
bound complex to understand PL at $d\ge2\lambda$ is $h$QE$_2$, which 
has the largest binding energy $\Delta$, and significant excitation 
energy $E^*-E$.
The $h$QE is also a fairly strongly bound complex with large excitation 
energy, but the charge neutral ``anyon exciton'' suggested by Rashba et 
al.\cite{rashba} is not bound.
It will be shown in a subsequent publication,\cite{bilayer-pl} the 
$h$QE$_2$ complex has a significant PL oscillator strength, while 
neither $h$QE nor $h$QE$_3$ are radiative.
Finally, the radiative excitonic state (charge neutral $e$--$h$ pair 
weakly coupled to the 2DEG) breaks apart at $d>2\lambda$.

\section{Conclusion}
\label{secVIII}

Using exact numerical diagonalization, we have studied energy spectra
of a 2DEG in the FQH regime interacting with an optically injected 
valence band hole confined to a parallel 2D layer.
Depending on the separation $d$ between the electron and hole layers, 
different response of the 2DEG to the hole has been found.
At $d$ smaller than a magnetic length $\lambda$, the hole binds one 
or two electrons to form neutral ($X$) or charged ($X^-$) excitons.
The $X$'s are weakly coupled to the 2DEG, and the $X^-$'s with the 
remaining electrons form a two-component fluid with Laughlin correlations.
One or two of the QH excitations of this fluid can bind to an $X^-$ 
to form a $X^-$QH$_n$ complex.
The PL spectrum at small $d$ depends on the lifetimes and binding 
energies of the $X$ and $X^-$ states, rather than on the original 
correlations of the 2DEG.
No anomaly occurs in PL at the Laughlin filling factor $\nu={1\over3}$, 
at which the FQH effect is observed in transport experiments.

At $d$ larger than about $2\lambda$, the Coulomb potential of the 
distant hole becomes too weak and its range becomes too large to 
bind individual electrons and form the $X$ or $X^-$ states.
Instead, fractionally charged excitons $h$QE$_n$ are formed, 
consisting of one or two Laughlin QE's bound to the hole.
Different $h$QE$_n$ complexes have different optical properties
\cite{bilayer-pl} (recombination lifetimes and energies), and which 
of them occur depends critically on whether QE's are present in 
the 2DEG.
Hence, discontinuities occur in the PL spectrum at $\nu={1\over3}$. 

The crossover between the ``integrally'' and ``fractionally'' charged 
exciton phases in an $e$--$h$ system can be viewed as a change in the 
response of a 2DEG to a more general perturbation potential $V_{UD}$ 
defined in terms of its characteristic energy ($U$) and length ($D$) 
scales. 
An analogous transition will occur in other similar systems, in which 
the 2DEG is perturbed by a charged impurity\cite{rezayi1,fox} or an 
electrode.
However, a difference between the response to negatively and positively 
charged probes is expected because of very different QE--QE and QH--QH 
interactions at short range.

Our results invalidate two suggestive concepts proposed to understand
the numerical $Ne$--$h$ spectra and the observed PL of a 2DEG.
First, in contrast with the works of Wang et al.,\cite{wang} and 
Apalkov and Rashba,\cite{apalkov} we have shown that the ``dressed 
exciton'' states with finite momentum ($k\ne0$) do not occur in the 
low energy spectra of $e$--$h$ systems at small $d$.
The coupling of $k\ne0$ excitons to the 2DEG is too strong to be 
treated perturbatively, and does more than renormalization of the 
exciton mass.
Rather, it causes instability of $k\ne0$ excitons and formation of 
charged excitons $X^-$.
And second, we have shown in contrast with the work of Rashba and 
Portnoi,\cite{rashba} that the charge-neutral ``anyon excitons'' 
$h$QE$_3$ are not stable at any value of $d$ (they are also 
non-radiative\cite{bilayer-pl}).

\section*{Acknowledgment}
\label{secIX}
The authors acknowledge partial support by the Materials Research Program 
of Basic Energy Sciences, US Department of Energy, and thank K. S. Yi 
(Pusan National University, Korea) who participated in the early stages 
of this study, and P. Hawrylak (National Research Council, Canada) and 
M. Potemski (High Magnetic Field Laboratory, Grenoble, France) for 
helpful discussions.
AW acknowledges partial support from the Polish State Committee 
for Scientific Research (KBN) grant 2P03B11118.

\end{document}